\documentclass[traditabstract]{aa}
\usepackage{graphicx}
\usepackage{graphicx}
\usepackage{subcaption}
\usepackage{amsmath}
\newcommand{\Msun}{M\ensuremath{_\odot}\,}
\usepackage{natbib}
\usepackage{color}
\usepackage[english]{babel}
\bibpunct{(}{)}{;}{a}{}{,} 

\begin{document}

\title{
Revisiting long-standing puzzles of the Milky Way: the Sun and its vicinity as typical outer disk chemical evolution
}

\titlerunning{The chemical evolution of the solar vicinity}

\author{M. Haywood\inst{1,3}, O. Snaith\inst{2}, M.~D. Lehnert\inst{3}, P. Di Matteo\inst{1,3}, S. Khoperskov\inst{1,4}}

\authorrunning{Haywood et al.}

\institute{GEPI, Observatoire de Paris, PSL Research University, CNRS, Sorbonne Paris Cit\'e, 5 place Jules Janssen, 92190 Meudon, France\\
\email{Misha.Haywood@obspm.fr}
\and 
School of Physics, Korea Institute for Advanced Study, 85 Hoegiro, 
Dongdaemun-gu, Seoul 02455, Republic of Korea  
\and
Sorbonne Universit\'{e}, CNRS UMR 7095, Institut d'Astrophysique de Paris, 98bis bd Arago, 75014 Paris, France
\and
Max Planck Institute for Extraterrestrial Physics, 85741 Garching, Germany
}

\date{Accepted, Received}

\abstract{
We present a scenario of the chemical enrichment of the solar neighborhood
that solves the G-dwarf problem by taking into account constraints on a
larger scale.  We argue that the Milky Way disk within 10~kpc has been
enriched to solar metallicity by a massive stellar population: the
thick disk, which itself formed from a massive turbulent gaseous disk.
While the inner disk, R$\lesssim$6 kpc, continued this enrichment after a
quenching phase (7-10~Gyr), at larger distances radial flows
of gas diluted the metals left by the thick disk formation at a time we
estimate to be 7-8 Gyr ago, thus partitioning the disk into an inner and outer region
characterized by different chemical evolutions.  
The key new consideration is that
the pre-enrichment provided by the thick disk is not related to the mass fraction of this
stellar population at the solar radius, as is classically assumed in
inside-out scenarios, but is actually related to the formation of the
entire massive thick disk, due to the vigorous
gas phase mixing that occurred during its formation.  Hence, the fact
that this population represents only 15-25\% of the local stellar surface
density today, or 5-10\% of the local volume density, is irrelevant for
``solving'' the G-dwarf problem. The only condition for this scenario to work is that the thick disk was
formed from a turbulent gaseous disk that permitted a homogeneous -- not
radially dependent -- distribution of metals, allowing the solar ring to
be enriched to solar metallicity. 
At the solar radius, the gas flowing from the outer disk combined with the
solar metallicity gas left over from thick disk formation, providing
the fuel necessary to form the thin disk at the correct metallicity to solve the G-dwarf problem.
  Chemical evolution at R$>$6~kpc, and in particular 
beyond the solar radius, can be reproduced with the same scheme.  
We suggest that the dilution,  occurring at the fringe of the thick disk, was
possibly triggered by the formation of the bar and the establishment
of the outer Lindblad resonance (OLR), enabling the inflow of 
metal poorer gas from the outer disk to R$\sim$ 6~kpc, presumably the
position of the OLR at this epoch, and at the same time isolating the inner disk from external influence. 
These results imply that the local metallicity distribution is not connected to the gas accretion 
history of the Milky Way.
Finally, we argue that the Sun is the result of the evolution typical of stars in the disk beyond $\sim$6~kpc (i.e.,
also undergoing dilution), and has none of the characteristics of
inner disk stars.
}

\maketitle

\section{Introduction}

Cold gas accretion \citep{dekel2006,woods2014, tillson2015}, which in
the last ten years has become the new paradigm describing how galaxies
acquire their gas, predicts that considerable gas accretion occurs
along a few dark matter filaments \citep{birnboim2003, keres2005,
ocvirk2008, keres2009, agertz2009, cornuault2018}, driving large amounts
of fuel in the inner parts of galaxies, permitting the early buildup 
of large disks \citep{genzel2006, toft2017, genzel2017}, and possibly
leading to the formation of large gas reservoirs  \citep{dave2012,
papovich2011, hopkins2014, suess2017}. Observations show 
that disks are indeed already massive at $z\sim1.5$, with roughly half
their stellar mass already in place for Milky Way-mass galaxies
\citep{muzzin2013, vandokkum2013, patel2013, papovich2015}, and perhaps
as much mass in molecular gas \citep{tacconi2013, dessauges2015,
saintonge2013, papovich2016}. 

Our Galaxy is compatible with this overall picture. In 
\cite{snaith2015}, \cite{haywood2016a} and \cite{haywood2018} we   show that the main chemical properties
of the inner Milky Way, R$\lessapprox$ 6~kpc (disk and bulge), are compatible
with a scheme where the gas has been accreted early
by our Galaxy. The substantial number of  low-metallicity dwarf stars that exist in the inner Milky Way (see the metallicity distribution function, MDF, in  \citealt{anders2014} or \citealt{haywood2018}) is in agreement 
with the predictions of a closed-box model where the
star formation history (SFH) has two predominate phases,  one corresponding to the growth of
the thick disk and the other to the growth of the thin disk. In
\cite{snaith2014, snaith2015} we  show that half of the disk stellar mass is
due to the thick disk. The thick disk stars that we see in the solar
vicinity are therefore the ``tip of the iceberg'' of a significantly more massive population, 
which, having a 
short scale length \citep{bensby2011,cheng2012,bovy2012b}, is mainly confined
in the inner Milky Way.  In standard
chemical evolution models  \citep[e.g.,][]{chiappini1997, colavitti2009,
marcon2010, minchev2014, kubryk2015}, the presence of such a huge
stellar population has a limited consequence on the evolution at the
solar circle because the Galactic disk is conceived as being made of independent rings whose
evolution is usually not connected to the others in order to reproduce
the inside-out paradigm of galaxy evolution. Hence, in these models the chemical
evolution of the thick disk seen at the solar vicinity is simply dictated 
by the evolution of the fractional mass of this population {\it \emph{at}} the
solar vicinity, while the evolution of the stellar populations in
the inner disk has no impact.  This independence with radius
of standard chemical evolution models has been mitigated in the last decade by
allowing for an unconstrained amount of radial migration of the stars,
in effect allowing yet other set of free parameters within models to
fit the dispersion and the mode of the metallicity distributions within
the Milky Way \citep[e.g.,][]{minchev2013,kubryk2015,loebman2016,toyouchi2018}.

In \cite{haywood2013,haywood2015, haywood2018}, we argued that there is
good evidence that the formation of the thick disk is not inside-out.
This is also what is observed in APOGEE: the chemical track of alpha-rich
stars in the [Fe/H]-[$\alpha$/Fe] plane is independent of the distance
to the Galactic center \citep{hayden2015}.  We advocated that the conditions that must have
prevailed in the interstellar medium (ISM) of disks at redshift greater
than 2, allowing for strong turbulence and feedback from vigorous star
formation \citep[see also][]{lehnert2014}, must have favored large mixing
of chemical species, explaining the lack of evidence of an inside-out
formation of the thick disk. This has important consequences on how we
see the chemical evolution of the disk at the solar vicinity, and in
particular on what is known as the G-dwarf problem.

The G-dwarf problem \citep{vandenbergh1962,pagel1975},  one of the longest
standing problems in galactic astrophysics,  is the recognition
that {\it \emph{local}} data offer no simple explanation of how the Galaxy reached
the metallicity above which most stars are found at the solar vicinity
(or [Fe/H]$\sim$-0.2). 

If most of the gas had been in the disk at 
early times, thus actively forming stars (as suggested by the picture outlined above), the number of stars 
at [Fe/H]$\leq$-0.2 necessary to increase the metallicity of this large pool
would have had to be a sizable fraction of the present local stellar
density; instead,  there are (at least in the solar vicinity)  only a relatively small number of these stars.
The local fraction of thick disk stars seems, on the contrary, to imply models where the disk would have been 
parsimoniously supplied with  gas (the gas infall models, see references above), again at variance with the general picture sketched out
at the beginning of this introduction. 

In the present study, we explore the simple idea that if the formation of
the thick disk is a global process (i.e., not inside-out), the enrichment
it provides cannot be accounted for in proportion to its local mass fraction,
but that it results from the chemical evolution of an entire massive 
population (a few
10$^{10}$\Msun) of the inner disk. In this scheme, the solar ring, although at the outskirts of the
thick disk, may have been enriched by this massive
 stellar population of the thick disk, due to the efficient mixing
within the  ISM
that prevailed at this epoch, thus solving the long standing
G-dwarf problem.  By describing the evolution of the outer disk,  which 
 includes the solar vicinity (see Section \ref{sec:sun}), this work complements
our  investigation of the evolution of the inner
disk of the Galaxy, R$<$ 6~kpc \citep{haywood2018}.  The outline of the
paper is as follows. In   Section 2 we first start by revisiting
the constraints provided by the [Fe/H]-[$\alpha$/Fe], age-[$\alpha$/Fe]
distributions of stars and the radial metallicity gradient.  In Section
\ref{sec:solar} we explain how our new scenario is applicable to
the evolution of stars at the solar ring, given the constraints advocated in
the previous section. In Section \ref{sec:outerdisk} we generalize our new
picture to the outer disk.  In Section \ref{sec:discussion} we discuss some
important issues. We present our conclusions in Section \ref{sec:conclusion}.

\section{Deconstructing the local chemical patterns: the [Fe/H] -
[$\alpha$/Fe] and age - [$\alpha$/Fe] planes}\label{sec:data}

At the solar vicinity most of the stars have a metallicity [Fe/H]$>$-0.2.
Understanding how the Milky Way ISM was locally enriched to [Fe/H]$\sim$ $-$0.2 
implies that we correctly interpret the two
sequences that have stars below this limit of $-$0.2:  the low  and
high alpha-abundances.  
These two sequences are
dominant in the inner and outer disks, respectively \citep{hayden2015}.  

We now review the constraints
offered by these two sequences when combined with stellar ages. We
also refer the reader to the study of \cite{buder2018} and the GALAH survey.

\subsection{Inner disk sequence: a temporal sequence}\label{upper}

\begin{figure}
\includegraphics[width=9.cm]{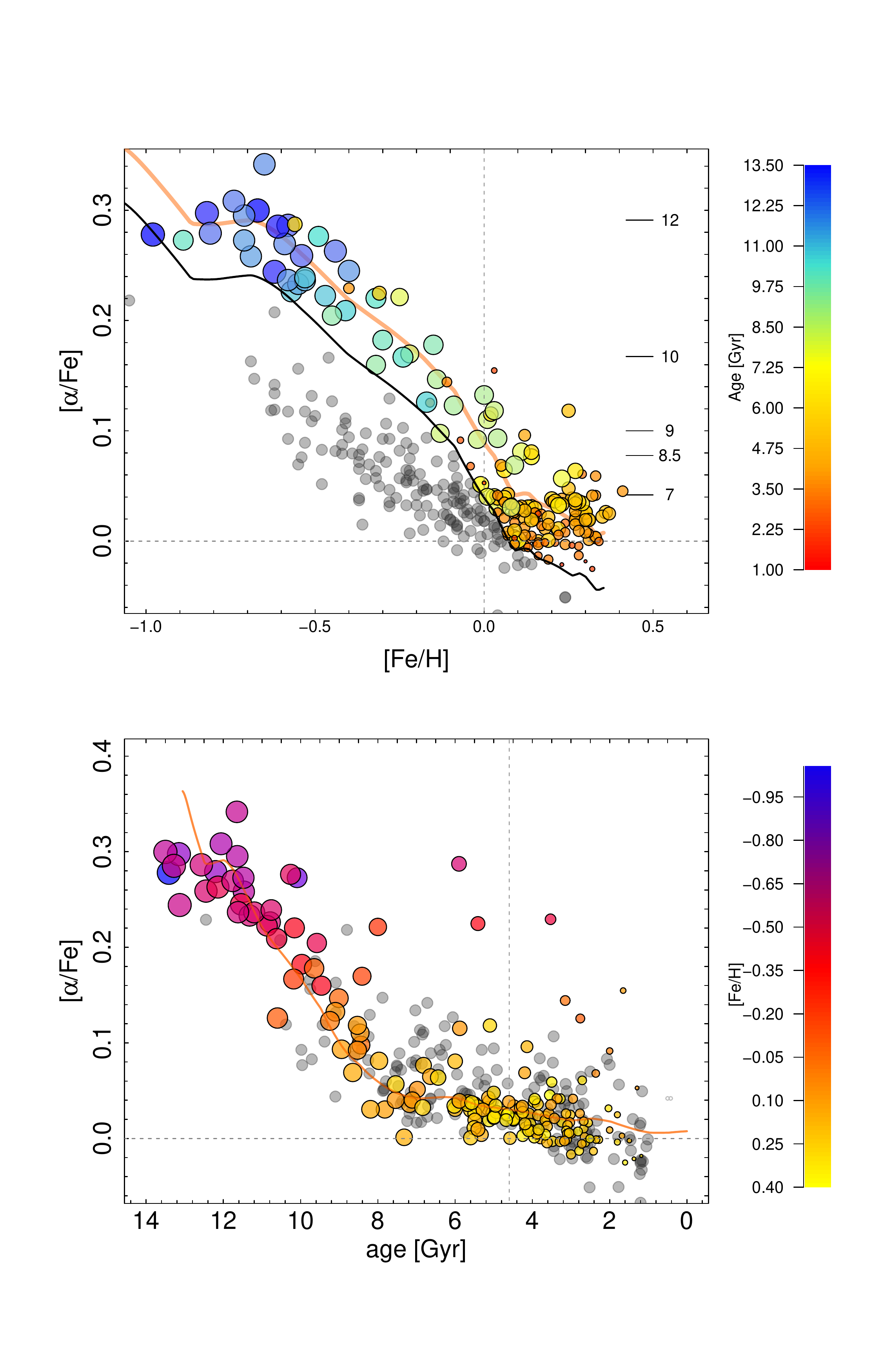}
\caption{
        \textit{Top:} Distribution of stars in the [Fe/H]-[$\alpha$/Fe] plane.  The color
and size of the symbols indicate the age of the stars, as shown by
the vertical colorbar on the right of the panel. The orange curve is our model 
        for the inner disk 
\citep{haywood2018, snaith2015}.  Colored points are selected
to be above the model curve shifted vertically by $-$0.05 dex.  The ticks
and numbers on the right side of the plot indicate the age of the
        model at different alpha-abundances.  \textit{Bottom:} Distribution of
stars in the age-[$\alpha$/Fe] plane, with stars selected as in the plot above.
The color and size of the points now indicate the metallicity
of the stars (right vertical colorbar).  }
\label{fig:alphafeh}
\end{figure}

Figure \ref{fig:alphafeh} shows the [Fe/H]-[$\alpha$/Fe] and
age-[$\alpha$/Fe] distributions of stars from \cite{adibekyan2012}
with ages from \cite{haywood2013}.  
The atmospheric parameters and 
chemical compositions are taken from \cite{adibekyan2012}, a sample of
nearby targets observed for the purposes of searching extrasolar planets.
This produced a sample of 1111 stars with temperatures, elemental
abundances, stellar velocities, and associated errors. As described in
\cite{haywood2013}, the selection of stars with age determinations with
adequate accuracy means that we had to severely prune the original
sample to only 363 stars. 
See \cite{haywood2013, haywood2015} for detailed explanations on how the
ages were derived and the expected accuracies.
Since our results are not derived from
stellar densities along the age-[$\alpha$/Fe] relation or in the [Fe/H]-[$\alpha$/Fe] plane, 
there should be no bias due to the volume definition
of the sample. Our sample is as complete as other larger surveys within
the range -1$<$[Fe/H]$<$0.5 in the [Fe/H]-[$\alpha$/Fe] plane.

In a recent
study, \cite{silva2018} claims, on
the basis of their asteroseimologic ages, that a tight
relation between age and [$\alpha$/Fe] or [Fe/H] does not exist during the
formation of the thick disk, in contradiction with
our results. 
However, if there are no relations between age-[Fe/H] and age-[$\alpha$/Fe],
it means that on the high-$\alpha$ sequence a star of
a given low metallicity and  high alpha-abundance can be younger than
another star with higher metallicity and lower alpha-abundance. Because
the  high-$\alpha$ sequence (in the [Fe/H]-[$\alpha$/Fe] plane) is continuous and has relatively small scatter, it would
be extremely difficult to explain how the chemical evolution proceeded
to maintain the underlying complexity in the age-[Fe/H]-[$\alpha$/Fe]
space while maintaining such a tight relation in the [Fe/H]-[$\alpha$/Fe] plane. The natural evolution
of all chemical evolution models is to decrease relative alpha-abundances
while increasing metallicity with time. While heterogenous chemical
evolution may exist, we see no reason why it would have conspired to give
a tight high-$\alpha$ sequence.  Second, [$\alpha$/Fe]-kinematics and
[$\alpha$/Fe]-structural parameters do show conspicuous correlations:
[$\alpha$/Fe] abundances correlate well on the
high-$\alpha$ sequence with an increase in the velocity dispersions
or orbital parameters \citep{haywood2013,bensby2014}. We also do not see  how
these correlations could exist at all if [$\alpha$/Fe] was not correlated with age, or was
only loosely correlated. Likewise, the increase in scale heights with
[$\alpha$/Fe] abundance \citep{bovy2012b} would simply not exist
without a relation linking age and [$\alpha$/Fe]. Moreover,
given that [$\alpha$/Fe] derived from SEGUE
are based on low signal-to-noise, low-dispersion spectra,
the underlying correlation between age and [$\alpha$/Fe] abundance must be
rather strong to still be  visible in the data \citep{bovy2012b}. Our final argument comes
from \cite{silva2018} themselves and their Figs. 3 and 10. These plots show that,
for stars older than 5 Gyr, the relative uncertainties in the ages they estimate are mostly greater than 30\%,
which for a 10 Gyr object is $\pm$3 Gyr. This is reflected in their
Fig. 10. Given these large uncertainties, any tight age-[$\alpha$/Fe] relation will be hidden in the 
observational errors.

The orange curve in Fig.~\ref{fig:alphafeh} represents our model describing the chemical evolution of the
inner disk and bulge \citep[see][]{snaith2015,haywood2018}.  We select
high-$\alpha$ stars in Fig.~\ref{fig:alphafeh}a by imposing that they must have an [$\alpha$/Fe] abundance
higher than the model shifted by -0.05 dex.  In \cite{haywood2018},
we show that the high-$\alpha$ sequence represents the evolution
of the inner disk and bulge and can be described by a model where
most of the gas has been accreted rapidly onto the disk.  The model curve
on the bottom plot shows two segments, representing the thick and
thin (inner) disks \citep[see][for a discussion]{haywood2013}. We
showed in \cite{haywood2016a} that the two phases are separated by
a quenching episode that occurred approximately 8-10 Gyrs ago.  The $\alpha$-rich
sequence is a temporal sequence: alpha-abundance and metallicities
are closely correlated with age \citep{haywood2013}, as can be seen in the bottom plot, 
even though the stars in the sample originate from different radii. 
This is why on the high-$\alpha$ sequence, mono-abundance
populations are also mono-age populations.

According to the studies of \cite{snaith2015}, \cite{haywood2015} and \cite{haywood2018}, this strong correlation 
between ages, alpha-abundances, and metallicities is due to
the closed-box type evolution of the inner disk and bulge, described globally by a very homogeneous
chemical evolution. It is apparent that chemical evolution proceeds
along this sequence. The best evidence for this is that the low-$\alpha$
sequence stars are essentially absent from the inner disk, except at high metallicity, which 
represents the evolution of the inner thin disk. Low-$\alpha$
stars are seen at a radius of R$\sim$5~kpc, but this is the tail of
a distribution that dominates the outer disk \citep[see][]{hayden2015}.
This spatial dichotomy invalidates scenarios trying to explain the high-$\alpha$ stars as the parent generation of low-$\alpha$ sequence stars
\citep[e.g.,][their Fig. 17]{schonrich2009,nidever2014}.  The crucial point for the rest
of the article, though, as argued in \cite{haywood2015, haywood2016a},
is that the homogeneous chemical evolution during the thick disk phase
implies that no inside-out or radially dependent formation occurred
for this stellar population. Another important point to take from
these two plots (see below) is that the thick disk phase reaches solar
metallicity, as is already known \citep[e.g.,][]{bensby2007}. The seven
stars with ages between 8 and 9 Gyr and [$\alpha$/Fe]$>$0.05 in the
top plot have a mean metallicity of $-$0.015 dex (1$\sigma$ dispersion of
0.10 dex).  We note that this metallicity is higher than the metallicity
of the oldest thin disk stars in the solar vicinity, implying that some
amount of dilution must have occurred (see below).

\subsection{Outer disk sequence: a dilution sequence}

\begin{figure}
\includegraphics[width=9.cm]{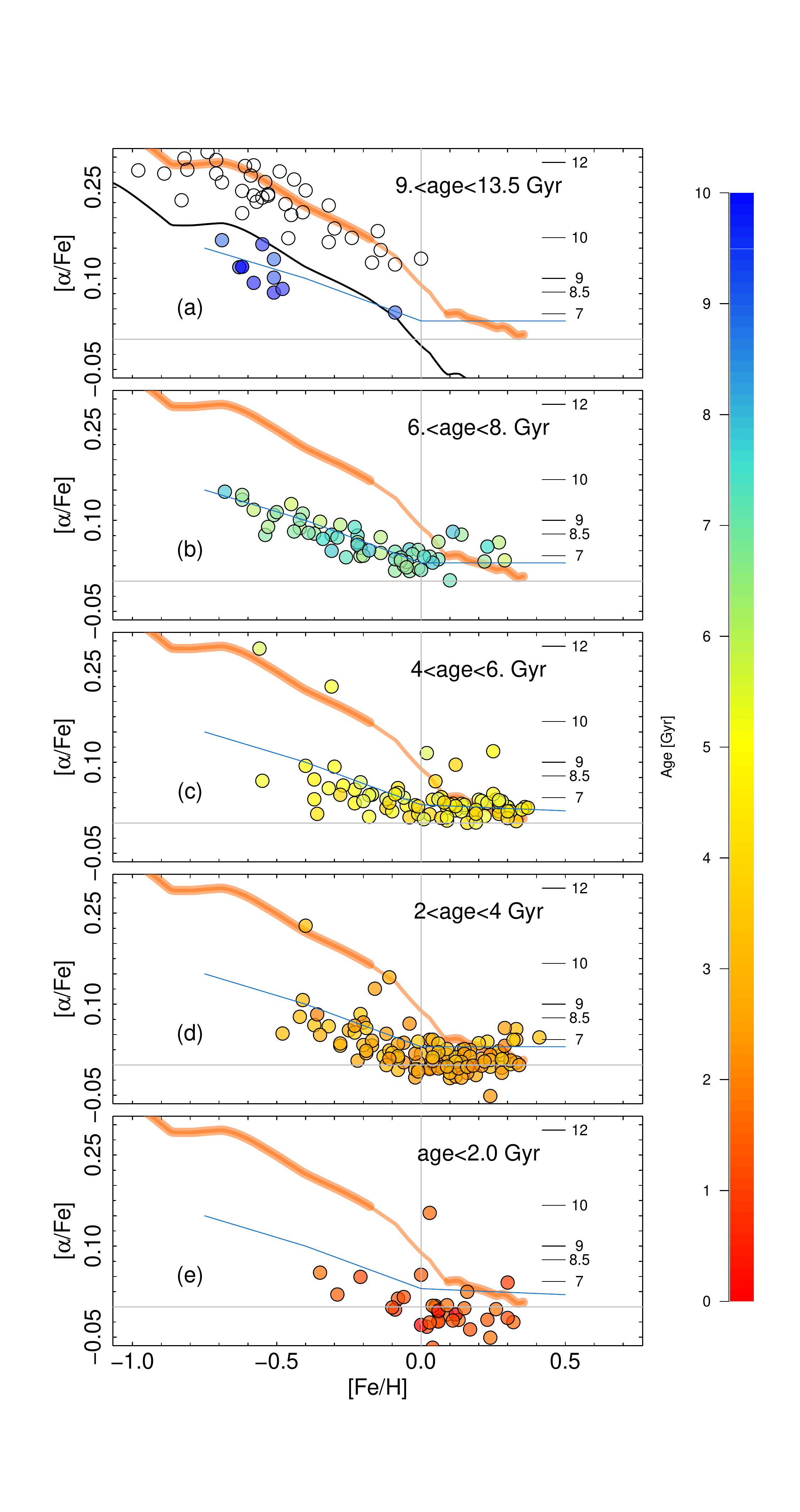}
\caption{
Distribution of stars in the [Fe/H]-[$\alpha$/Fe] plane as a function of age range (top to
bottom plots). The orange curve is our model for the inner disk. Ages from 12 to 7 Gyr
along this sequence are also indicated.
The blue curve is the same in all plots and serves as a guide to show the shift 
of the distribution with age. The parallel sequences of different ages illustrate that 
chemical evolution in the solar vicinity (the OLR region) and beyond proceeded along
evolutionary paths that link these sequences as age-metallicity sequences with the two sequences
being of different age. }
\label{fig:alphafeh-age}
\end{figure}

Figure \ref{fig:alphafeh-age} shows our sample sliced in different age
intervals \citep[see also ][their Fig. 22]{buder2018}.  The orange curve
represents our model track describing the evolution of the inner disk and
bulge, as in Fig. \ref{fig:alphafeh}.  This figure illustrates that the
outer disk sequence is clearly stratified in age and metallicity, with
older stars at higher alphas for a given metallicity,  as is already
known from the age-[$\alpha$/Fe] relation.  Mono-abundance populations in this sequence 
are not mono-age populations because stars of a given age cover a large
range in metallicity.  The missing parameter needed to single out a
mono-age population on the outer disk sequence is the birth radius of the stars, as we discuss below.

The sequences of coeval stars in Fig. \ref{fig:alphafeh-age} therefore
suggest that the evolution is not from the most metal-poor stars, at [Fe/H]$\sim -$0.7, to
the most metal-rich, at [Fe/H]$\sim$+0.5,   as  is  known, also from the age-metallicity
relation \citep[see, e.g.,][]{edvardsson1993,haywood2006,casagrande2011}.
Therefore, the low-$\alpha$ sequence is not a temporal
sequence: while the oldest stars are also the most metal-poor (plot a,
see also \citet{haywood2013,buder2018}), the most metal-rich are found
at all ages below 8 Gyr.  Instead, in the sequence of oldest stars, at ages between 9 and 8 Gyr, we see the first
generation of stars that started to form at different metallicities in
the thin disk, with dilution in stars with higher alpha-abundances. 
The youngest sequence (age$<$ 2Gyr, plot e) is the end point
of evolutions that start from the oldest sequence (plot b).
Because the dynamical properties
of stars at the lowest metallicities in the solar vicinity suggest
they come from the outer disk \citep{haywood2008}, and because APOGEE
observed the same type of stars in situ in the outer disk, it follows
that each sequence in Fig.\ref{fig:alphafeh-age} is also a sequence
dependent on the birth radius of the stars. Interpreting the data this
way suggests that the thin disk started to form stars with decreasing
metallicity at increasing radius. Hence it is natural to interpret the
low-$\alpha$ sequence observed within the solar vicinity as a composite of
chemical tracks, each describing an evolution at a given radius, slightly
increasing in metallicity and decreasing in alpha-abundance as a function
of time.  We are able to observe this complexity in the solar vicinity, due to
the dynamical wandering of stars born at all radii (i.e., the amplitude of their
radial oscillation around their guiding centers).  For the solar vicinity,
such a track would show a mean evolution of metallicity as a function
of age limited to a range from about $-$0.2 to about 0.1-0.2, as can be
measured on the age-metallicity relation \cite[see][]{haywood2006,haywood2013}.
The evolution at other radii can be conceived similarly, the only
difference being the initial metallicity and alpha-abundance, which
at R$>$6 kpc are both a function of the distance to the Galactic center.

\begin{figure}
\includegraphics[width=10.cm]{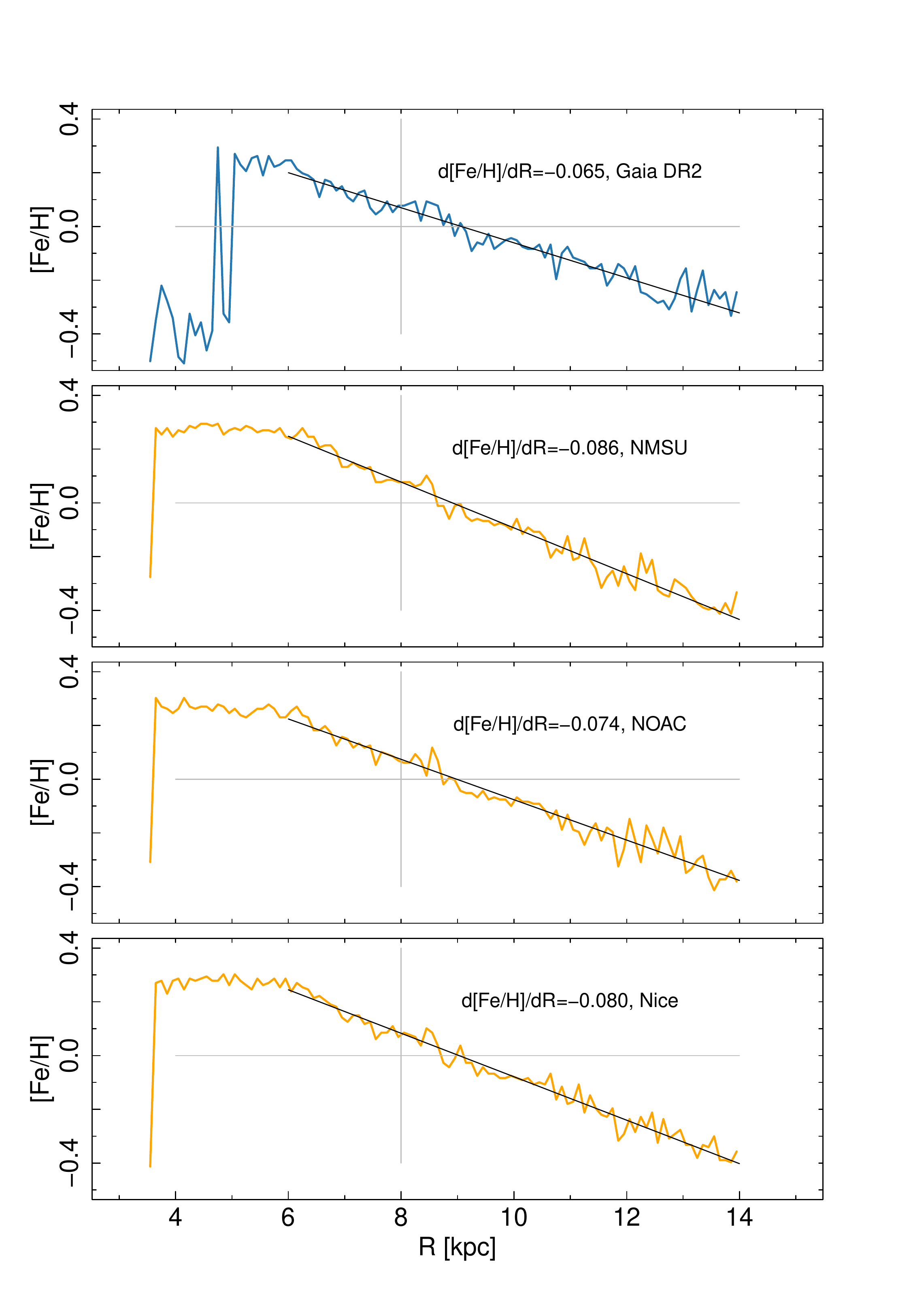}
\caption{Metallicity of the peak of distribution of metallicities of
low-$\alpha$ stars in APOGEE as a function of R for different distance
estimates: Gaia (top) and then NMSU, NAOC, and Nice in the next three panels
from top to bottom (as indicated, along with the radial gradient estimate,
in the legend of each panel).  At R$<$6 kpc, the gradient is flat in
all estimates (inner disk).}
\label{fig:fehgradient} \end{figure}

We can quantify how these trends change with radius and more
generally we can determine  the metallicity profile of the thin disk.
Figure~\ref{fig:fehgradient} shows the metallicity of the peak of the
distribution of the low-$\alpha$ sequence as a function of R using the data from the DR14 \citep{abolfathi2018}
of the near-infrared high-resolution (R$\sim$22500) spectroscopic survey APOGEE \citep{majewski2017}. The distances for the stars are taken from the 
parallax estimates of the Gaia DR2 (top), selecting stars with less than
20\% relative error on parallaxes, and distance estimates of 
\citet[][labeled NMSU]{holtzman2018}, \citet[][labeled NOAC]{wang2016} and
\citet[][labeled Nice]{schultheis2014}.  
We use all stars in APOGEE DR14 that have a distance estimate, a
signal-to-noise ratio higher than 50, effective temperatures lower than
5250~K, and log~g$<$3.8.

All plots except the one with
Gaia parallaxes, which lacks data for a number of APOGEE stars in zones
of high extinction and crowding, show two distinct regimes: the first 
at R$<$6~kpc shows a flat gradient, and the second above
this limit that shows a gradient between 0.065 to 0.086~dex.kpc$^{-1}$.
These gradients are steeper than those found by \cite{hayden2015},
the main reason being the metallicity estimator used in each case (value of
the peak in our case, and the mean in their case), and there is a
difference of $\approx$0.15 dex between the two estimates.  We note that the
absolute values of the metallicities may slightly overestimate the real
metallicities of the population because giant stars are biased towards
younger ages \citep{bovy2014}.

The break in the metallicity profile of these two regions reflects the change in nature
of their chemical evolution: the evolution within R$\sim$6~kpc is
closed-box-like \citep{haywood2018}, while the evolution beyond R$\sim$6
kpc has an evolution that was significantly impacted by dilution of
the enriched gas from which it formed, as we describe below.  In the
outer disk, at R$>$6~kpc, the gradients show at what metallicity most of
the stars formed, ranging from almost +0.1 at solar radius to $-$0.4 at
14~kpc, illustrating the shift in the evolution as a function of radius,
with chemical evolution starting at decreasing lower initial metallicity
towards the outer disk, and having formed the majority of its stars also
at decreasing metallicities.

\section{Solar vicinity as the prototypical example of the outer
disk evolution}\label{sec:solar}

\begin{figure}
\includegraphics[trim=10 140 10 10,clip,width=9.5cm]{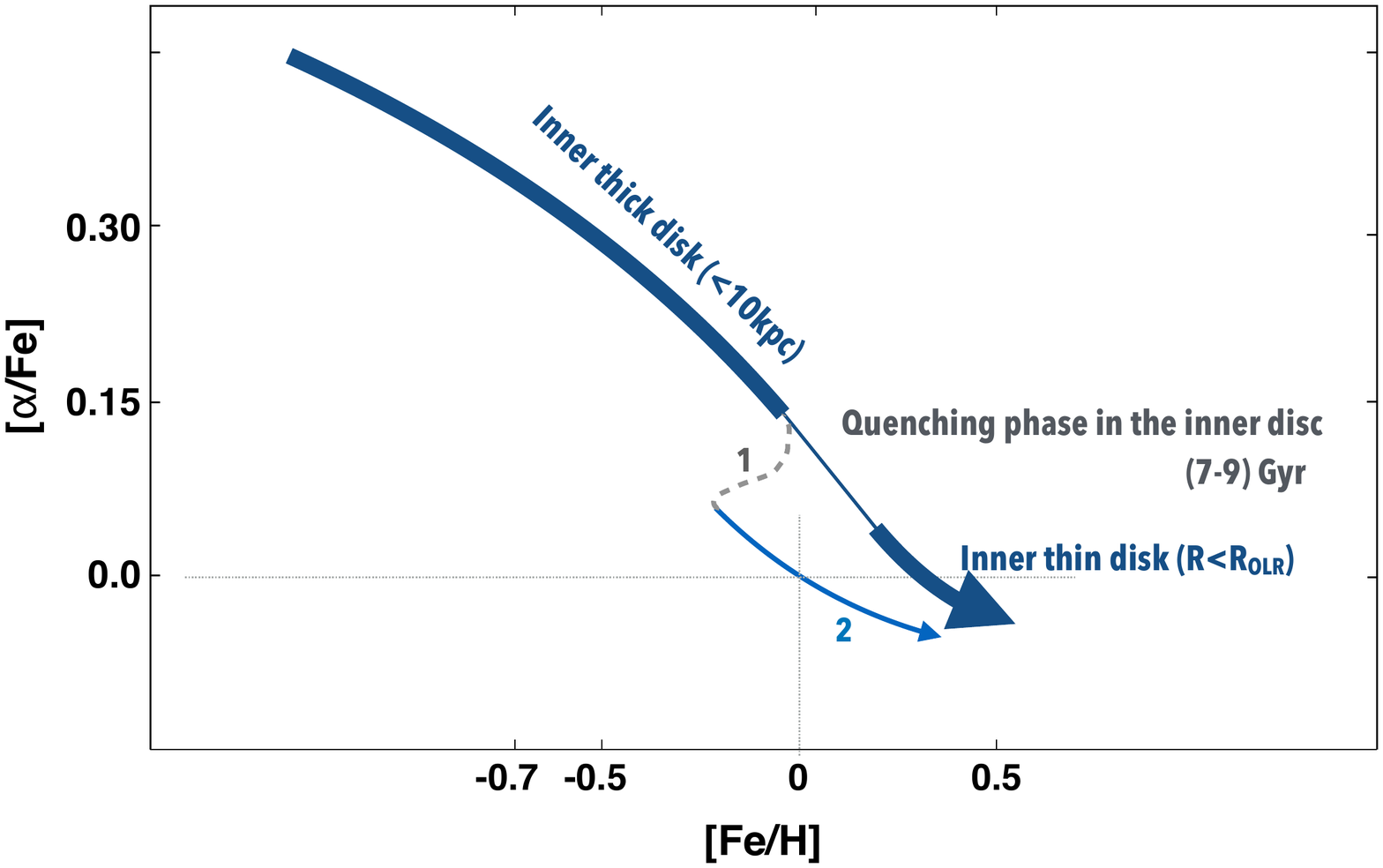}
\caption{Schematic representations of the chemical evolution at the solar
radius in the [Fe/H]-[$\alpha$/Fe] plane according to the description
presented in Section \ref{sec:solar}.  The blue thin curve with the arrow
(2) represents the chemical track for the solar ring evolution after an
episode of dilution that occurred at the end of the thick disk formation
(1).}
\label{fig:sketchsun}
\end{figure}

\subsection{Hints from the solar vicinity chemical patterns}

There are objective facts suggesting that dilution is needed to explain the 
chemical evolution at the solar radius, unrelated to the G-dwarf problem.
As mentioned above, there is evidence that the ISM at the end
of the thick disk phase ($\sim$8-9 Gyr ago) reached a near-solar
metallicity. At the same time, we know that the solar vicinity MDF peaks
at solar metallicity, with most of its stars being younger than 7-8 Gyr
(the Sun  being only 4.6 Gyr old at [Fe/H]=0.).  This tells us
that at the solar vicinity, solar metallicity was reached two times,
first about 8-9 Gyr ago at the end of the thick disk formation and
then $\sim$ 4.5 Gyr ago, implying a dilution episode between the two, as 
already suggested \citep[e.g.,][]{bensby2004}.

If we assume that these two points are representative of the metallicity
evolution at the solar ring, and given the constraints discussed above,
the track followed by chemical evolution at the solar ring must have been
similar to the chemical track illustrated in Fig. \ref{fig:sketchsun}.
First, the thick disk formation dominated the evolution (thick blue curve)
and at the end of the thick disk phase, a dilution occurred, lowering the
metallicity to $\sim$-0.2 dex (dashed curve), which is the metallicity of
the oldest thin disk stars locally \citep{haywood2006,haywood2008,casagrande2011}.
We note that the solar neighborhood  contains lower metallicity thin disk stars, but these are
believed to most likely come from the outer disk, as shown by their kinematic and orbital properties
\citep{haywood2008,bovy2012b}.
Then a moderate star formation rate (SFR: 2-3 $\rm M_{\odot}$.yr$^{-1}$) over
3-4 Gyr increased the metallicity by about +0.2 dex, reaching solar
metallicity for the second time.

At the solar radius, the thick disk surface density represents
about 12$\pm$4\% according to \cite{bland2016}. Measurements of
the surface densities of the stellar mono-abundance populations in
\cite{bovy2012a,bovy2012b} yield higher estimates, the differences being
largely a matter of how the thick disk is defined. For example, if it is
defined as stars having high alpha-abundance ($>$0.20 dex on the SEGUE
scale, more likely 0.15 dex with the stellar abundances used here), 
including stellar mono-abundance having scale heights larger than 400~pc,
and in this case using estimates from \cite{bovy2012a}  (as inferred from
their Fig. 2), it may represent as much as 25-30\%.  Adopting this upper limit, and given that Milky
Way mass galaxies are observed to contain a similar mass of stars and gas
at redshifts 1-1.5 \citep{tacconi2013, dessauges2015, saintonge2013,
papovich2016}, we can then assume that the total baryonic
mass (gas+stars)  at the end of the thick disk phase would represent 50-60\% of the
present-day surface density at the solar ring.  Thus, at the solar
ring, the gas left at the end of the thick disk phase would provide
an insufficient reservoir of gas to form the thin disk, and another
40-50\% of gas would be necessary to reach the observed present-day mass
surface density.  Again, we note that we are talking here of the
disk beyond $\sim$6~kpc, or beyond the possible initial position of the
OLR. Inside the OLR, the description given by the closed-box model as
given in \cite{haywood2018} shows that no supplementary replenishment
of gas is necessary to prolong its evolution up to the present time.

The solution to these two problems, namely the decrease in metallicity
of the ISM $\sim$8-9 Gyr ago and an insufficient amount of gas to form
the thin disk at the end of the thick disk phase at the solar vicinity,
implies a new supply of gas, and together provide the dilution
necessary to decrease the metallicity of the ISM to $\sim$-0.2 dex
after the thick disk formation, and the additional fuel to form the
thin disk at the solar vicinity.  The following back-of-the-envelope
calculation provides a hint to the possible origin of the incoming gas.
As already stated, it can be assumed that the surface density at the end of the thick disk phase 
was composed of  half  thick disk stars and  half 
 solar metallicity gas left over from the formation of the thick disk.
The amount of gas acquired to form the thin disk at the end of the formation
of the thick disk would then double this surface density to reach the present value.
If the incoming gas that mixed with the gas left over  from the thick disk
formation was near-pristine,  [Fe/H]$<$-2 dex, the metallicity of
the mixture from which the first thin disk stars started to form locally
would have been [Fe/H]$<$-0.5 dex, which is at least $-$0.3 dex too low
compared to what is observed.  Therefore, the inflowing gas must have
been significantly more metal rich.

The results of the APOGEE survey show that the metallicity of
giants at the largest distances in the outer disk is about $-$0.5
to $-$0.7.  The oldest (9-10 Gyr) low-$\alpha$ sequence  stars observed in the solar vicinity (see
Fig. \ref{fig:alphafeh-age}),
presumably of outer disk origin, have a similar metallicity.
This suggests that it is also the metallicity of the gas that was in
place beyond the thick disk, R$>$10~kpc, and that mixed with the
solar metallicity gas left over from the formation of the thick disk at
its outskirts.  If we fill out the 50-60\% of the gas that was missing
after the thick disk phase with gas at this metallicity, $\sim$ -0.6
dex, the thin disk phase would start its formation with a gas mixture
containing one-third of the gas coming from the thick disk at solar
metallicity and two-thirds coming from the outer disk with a metallicity
of $-$0.6.  This would decrease the metallicity of the gas from
about solar to $\approx$$-$0.2 to $-$0.3  dex.  Thus, if the fractions we
have estimated are correct, the fuel provided by the outer disk gas had
the right metallicity to form the thin disk stars at the solar radius
once it was mixed with the gas left over by the growth of the thick disk.

We note that in this scenario the decrease in the metallicity at
the end of the thick disk phase that we see at the solar vicinity is due
to the dilution of the ISM by gas from the outer disk (not infalling
pristine gas); it is not generated by the gap in the SFH described in
\cite{haywood2016a}, which occurred in the inner disk. In this regard,
it is also different from the results of \cite{chiappini1997}, who hypothesized that  the
decrease in metallicity was  generated by a gap in the
SFR  at the solar ring combined with a continuous infall of pristine gas.
\cite{haywood2018} argue that the data for the inner disk are compatible
with most of the gas being accreted very early in the Milky Way, with essentially
little or no  accretion after the thick disk phase within the extent of
the thick disk (R$<$10~kpc).

\subsection{Metal mixing in thick disks}

Is the assumption that the thick disk at 8-10~kpc had a similar chemical
evolution to that of  the inner regions realistic?  Observations of distant
galaxies as a function of redshift show that flat metallicity gradients
over a distance of $\sim$ 10 kpc are common, with a small spread from
galaxy to galaxy,  usually a few $\pm$0.01 dex/kpc \citep{stott2014,
wuyts2016, leethochawalit2016}.  The variation in gradients from galaxy
to galaxy has been attributed to feedback strength, following results
obtained by several groups simulating the formation of Milky Way-type galaxies.  For example, \cite{angles2014} show that simulations with no wind usually
generate steep gradients of metallicities, while Galactic outflows, by
allowing the redistribution of metal-enriched gas over large scales,
generate flat gradients, confirming similar results of other studies
\citep[e.g.,][]{gibson2013,ma2017}.  If the Milky Way thick disk  formed
from a turbulent thick layer of gas, as seems to be most probable, a
flat metallicity gradient would be a natural outcome in such a model, and we do expect
solar vicinity to have been enriched to the same level as the inner regions of
the thick disk (i.e., not inside-out).

\subsection{When did the dilution occur and why?}

We can constrain when the dilution episode occurred within a few Gyr:
it must have occurred after the thick disk reached solar metallicity,
about 8-9 Gyr ago, and significantly before  the
birth of the Sun (at least a few Gyr) so that the ISM had time to be enriched again to its
metallicity at the birth of the Sun 4.6 Gyr ago. It can be inferred then
that the dilution must have occurred $\sim$7-9 Gyr ago.  The dilution
could be the result of an accretion episode of gas by the Galaxy, but
various arguments suggest otherwise.  First, since the dilution seems to
have occurred within a relatively narrow timespan, this allows us to reject
the possibility that it was an effect of a long-timescale infall of gas.
Second, if the material that mixed with the gas left over by the formation
of the thick disk had a metallicity of $\sim$-0.6, as suggested above,
it is difficult to imagine that it was accreted directly from cold flows
and must have been in place before inflowing to the solar radius. The
metallicity of the intergalactic medium during this epoch was likely
much lower than this \citep{bergeron2002, simcoe2004, simcoe2011}.
Presumably, higher angular momentum, pristine, or very low-metallicity gas
was accreted in the outer parts of the Milky Way.  \citet{lehnert2014} suggest
that throughout the formation of the thick disk, the star formation
intensity of the Milky Way was well above the threshold for driving
outflows, which likely lead to gas in the outer disk being polluted by
inner disk gas,  raising its metallicity to $-$0.6, and fixing its
[$\alpha$/Fe] at $\sim$+0.15.  

Our ignorance of the gas accretion history, the metallicity of
the outer disk, and the rate and fate of metals that were expelled
by outflows during the formation of the thick disk means that we can
only provide speculative answers as to how this initial outer disk gas
composition may have been set.  We can make rough estimates to show
that this is not impossible, although there is no proof that it happened
this way. For instance, \cite{mackereth2017} show that the thick disk, or
more precisely the high-$\alpha$ population, and the low-$\alpha$
population at [Fe/H]$<$-0.2 dex have approximately the same surface
density locally. Assuming that they have scale lengths of 2 kpc and 4 kpc,
respectively, the outer disk at R$>$10kpc is $\sim$10 times less massive
than the thick disk.  The importance of outflows are similarly difficult
to estimate, but considering that we describe the evolution of the thick
disk as closely approximated by a closed-box, any metals lost via winds would
have to be limited, and in particular must not be significant enough to
substantially modify the overall metallicity distribution of the inner
disk \citep[unlike in the model of ][]{hartwick1976}.  If we assume that
no more than 5-10\% of the metals in the thick disk were expelled by
outflows at a metallicity between -0.6 and -0.2 dex, which corresponds
to the thick disk metallicity range at the maximum of the SFR, then a
fraction between 0.1$\times$0.005 and twice this amount of the thick disk
mass in metals may have polluted the outer disk. This assumes that all
metals eventually rain down to the Galactic plane before significant
star formation in the outer disk.  Diluted by a component roughly
ten times less massive than the thick disk, the metals would raise the
metallicity to a value similar to what it was in the thick disk, or -0.6
dex. These estimates are clearly very rough and neglect a number of
factors that may be significant.  For example, the amount of gas expelled
from the thick disk may  have been  smaller and still have provided
similar enrichment if the infall of more pristine gas from the Galactic halo
occurred over a long timescale, in which case the pool of gas receiving
the metals would have been even smaller and thus less diluted.  The gas
could then fall back onto the disk through a mechanism such as Galactic
fountains \citep{shapiro1976,bregman1980,marinacci2011,fraternali2017}.

How and when did this gas   mix with the gas left by the thick
disk formation?  The epoch of the formation of bars in galaxies of
the mass of the Milky Way predominately occurred $\sim$9-10 Gyr ago
\citep{sheth2008,melvin2014}.  If this is also the epoch of
the formation of the bar in our Galaxy, it must have been a time of
rearrangement of stars and gas due to the dynamical impact of the bar
potential.  If the formation of the bar is at the origin of the quenching
event, as suggested in \cite{haywood2016a}, and studied theoretically in
\cite{khoperskov2018}, it is tempting to also associate the dilution with
the impact of the bar on the gas.  If the quenching episode was triggered
by the formation of the bar, the outer Lindblad resonance must have been in
place about 9-10 Gyr ago.  Because of the clear difference of chemical
properties of the disk within and beyond $\sim$6~kpc observed today
(Fig.\ref{fig:fehgradient}), the OLR could well have been established
at this radius.  Estimates of the current position of the OLR vary
between 6-9~kpc \citep{dehnen2000}, 10-11~kpc \citep{liu2012}, and 7~kpc
\citep{monari2017}. Given that the OLR must have shifted to larger
radii as the bar pattern speed decreases, an initial position
at 6~kpc is thus plausible.
 At that time the thick disk, with its uniform metallicity and
 well-mixed gas,
extended to roughly 10~kpc.  Inside the OLR, the gas is driven from the
corotation to the OLR \citep{simkin1980,byrd1994,rautiainen2000}, helping
to maintain a zero gradient inherited from the thick disk formation
during the thin disk phase, as observed in Fig. \ref{fig:fehgradient}
(see next section). The action of the bar prevented the inner disk from
being subsequently diluted by radial flows. This effect is crucial to
the overall validity of our proposed scenario.

If the OLR was established at $\sim$6 kpc by the formation of the bar,
it may have taken $\sim$1-2 Gyr for the inflowing, metal-poor gas, which
was at the time at larger distances, to reach the OLR radius (inflowing at
a few km.s$^{-1}$).  Beyond  $\sim$6~kpc, the metal-poor gas mixed with
the solar metallicity gas with resulting mean metallicity decreasing
outwards.  We note that recent theoretical ideas about the impact of
flowing gas on the velocity dispersion in the ambient gas suggest that
this is a viable mechanism for increasing the amount of turbulence in
disks \citep{krumholz2018}. While we  do not yet understand the impact
of this increased turbulence may have on the star formation efficiency or
perhaps even in suppressing star formation in the outer disk, an increase
in the level of turbulence and the gas velocity dispersion would at the
very least lead to mixing between the ambient, leftover gas from the
formation of the thick disk and the less enriched gas from farther out
in the disk. The efficient mixing supports our picture in that it would lead
to a continuity in the metallicities that is observed in the outer disk where the
only dependence appears to be a radial one.
After the formation of the bar and the establishment of the OLR, the bar would slow down and 
the OLR would be displaced to a larger radius, but we would expect its effect as a barrier 
maintaining the inner and outer disks to continue, as commented in \cite{halle2015,halle2018}.

\subsection{Model for the solar vicinity}

\begin{figure}
\includegraphics[trim=0 0 0 0,clip,width=9cm]{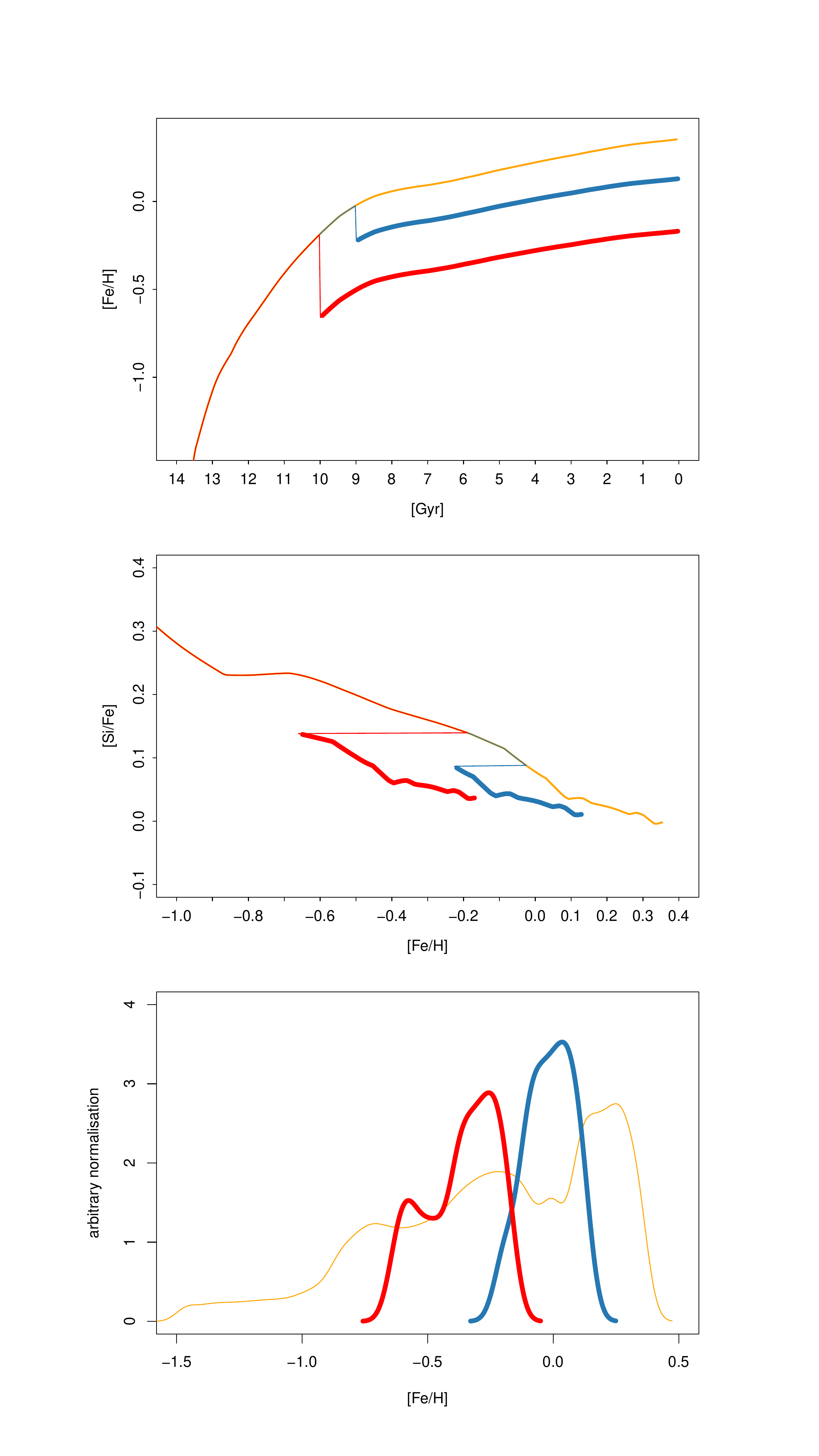}
\caption{Outer disk chemical evolution models according to the scheme explained in Sects.~\ref{sec:solar} and \ref{sec:outerdisk}, 
describing the evolution at the solar radius (blue curves) and farther out, starting from a metallicity of -0.6 dex.
In all plots, the orange curve is the closed-box chemical evolution model developed in \cite{snaith2015} and \cite{haywood2018}.
In the bottom plot, only the MDFs for stars younger than 
9 Gyr (blue curve) and 10 Gyr (red curve) are shown. Both MDFs have been smoothed by a Gaussian kernel of dispersion 0.04~dex.}
\label{fig:svodmodel}
\end{figure}

Figure \ref{fig:svodmodel} illustrates a model (blue curves) representing the solar vicinity chemical evolution according to the scheme presented above. 
The model is based on the closed-box model (shown as the orange curve on the plot) described in \cite{snaith2015} 
and which was shown to be valid to describe the whole inner disk and bulge \citep{haywood2018}.
The basic ingredients of the model are given in \cite{snaith2015}, together with a description of 
its main assumptions. 
The model for the solar metallicity first follows the closed-box chemical track from early times to 9 Gyrs ago, forming the
thick disk and reaching solar metallicity. 
An instantaneous dilution is then introduced at 9 Gyr and the metallicity is decreased from about solar to -0.2 dex. 
 The SFH of the model is obtained,  as in \cite{snaith2015}, by fitting the age-[Si/Fe] data of solar metallicity 
thin disk stars. The metallicity evolution and ([Fe/H],[Si/Fe]) distribution are shown in the first two plots 
of Fig. \ref{fig:svodmodel}.
The bottom plot shows the MDF for stars created in the model after 9 Gyr, or after dilution. They represent the 
thin disk part of the evolution.
It shows a distribution very near to what is observed in the solar vicinity for stars of the thin disk
(or low-$\alpha$ stars). The real MDF  would also include a small percentage of stars of the thick disk.  
In our scenario, the relative local fraction of the two is not related by chemical evolution (only the global fraction is), 
but is only an effect of the relative density distribution of the thin and thick disks locally. The
relative density distribution is a consequence of the formation, and 
subsequently of the dynamical processes that fix their scale lengths and scale heights.
As can be noted, the local MDF is at a maximum near solar metallicity, where the inner disk MDF reaches a
minimum due to the quenching episode that is observed on the APOGEE data \citep{haywood2016a} and on the bulge data 
\citep{haywood2018}.

\section{Generalization of our scenario to the whole outer disk}\label{sec:outerdisk}

Figure~\ref{fig:fehgradient} shows that the solar ring is only a part
of the radial metallicity profile which remains similar with a decreasing
metallicity down to the limit of the sampling provided by the APOGEE data.
It is therefore tempting to suggest that the scheme designed for the solar
vicinity could be extended to larger distances from the Galactic center (but also to smaller distances as the plot shows that the solar ring,
at 8 kpc, does not lie on the border of the inner disk).
\subsection{Models for the outer disk}

The previous section describes a two-step process. First, we imagine that the pristine gas in the outer
regions of the Milky Way disk is polluted by the outflows generated during the most intense phase
of the formation of the thick disk (from about 10 to 12 Gyr ago), raising its metallicity to 
about -0.6 dex. Second, at the fringe of the thick disk, this gas then mixes with the solar metallicity gas left at the end of the
formation of the thick disk, as described for the solar ring in the previous section, 
increasing its metallicity still further relative to the outer disk.
The validity of this scenario concerning the energetics of the outflows, angular momentum 
of the inflowing gas has to be worked out, but is clearly beyond the scope of the present work
where,  for the sake of clarity, we  concentrate solely on the chemical evolution of the outer disk.

At the solar ring, we  see that the gas from the outer disk diluted the ISM
left over by formation of the thick disk (from about 0 to -0.2 dex) and
provided the necessary and sufficient additional gas supply to form the thin
disk, as observed in the solar vicinity. How and why the dilution occurred
must be determined, but the scheme designed for the solar vicinity can be
generalized assuming that at larger R, the relative ratio of enriched
gas from the thick disk and the more metal-poor gas from the outer disk
decreases, leaving a mixture of decreasing metallicity to fuel star
formation. As discussed in section \ref{sec:data}, the evolution at a
given radius can then be thought of as parallel sequences mirroring the
evolution of that of the solar vicinity, but starting at an initial,
lower metallicity.  The decrease of this initial metallicity with R is
also reflected in the metallicity gradient of Fig. \ref{fig:fehgradient}.

Figure~\ref{fig:svodmodel} shows a model (red curves) where the formation of the thin disk 
starts slightly earlier, or 10 Gyr, and from a metallicity of -0.6 dex. This model 
could represent the most distant disk stars observed by APOGEE, at $\sim$ 14-15~kpc from 
the Galactic center. The model is built in the same way as the solar vicinity model,
except that because it is  several kpc from the edge of the thick disk we view 
the metallicity as being mainly the result of the mixture of pristine gas accreted from
the halo and enriched gas ejected from the thick disk, i.e., the first step mentioned 
at the beginning of this subsection. We note that the initial abundances of the model are fabricated 
this way, but it must be clear that the chemical track of the model in the first Gyrs
corresponds to stars formed in the thick disk. It is the formation and evolution of the thick disk that provides chemical enrichment to the 
outer disk via outflows, with roughly the metal budget discussed in the previous section, 
but that we do not expect a significant number of stars of the thick disk to have formed
in the outer disk.

The resulting chemical tracks are visible in the first two plots
of Fig.~\ref{fig:svodmodel}, while the third plot shows the MDF.
Because the model is constrained to fit the age-[Si/Fe] relation of
\cite{haywood2013}, it must follow the knee that is visible in this
relation, and which can only be fitted by lowering the SFR at about 9
Gyr, producing the dip that appears in the MDF at about $-$0.5.  The two
models presented in Fig.~\ref{fig:svodmodel} are representative of the
evolution at two different radii in the outer disk. The only difference
between the two is the initial chemical abundances from which each model
starts to evolve. These two examples give us the premise from which we can
conceive the evolution of the entire outer disk from R$\sim$6 to 15~kpc,
and understand the evolution underlying the classical chemical abundance
plots observed in the solar vicinity.

\subsection{Sketch of the chemical trends}

\begin{figure}
\includegraphics[trim=30 140 -100 10,clip,width=10.55cm]{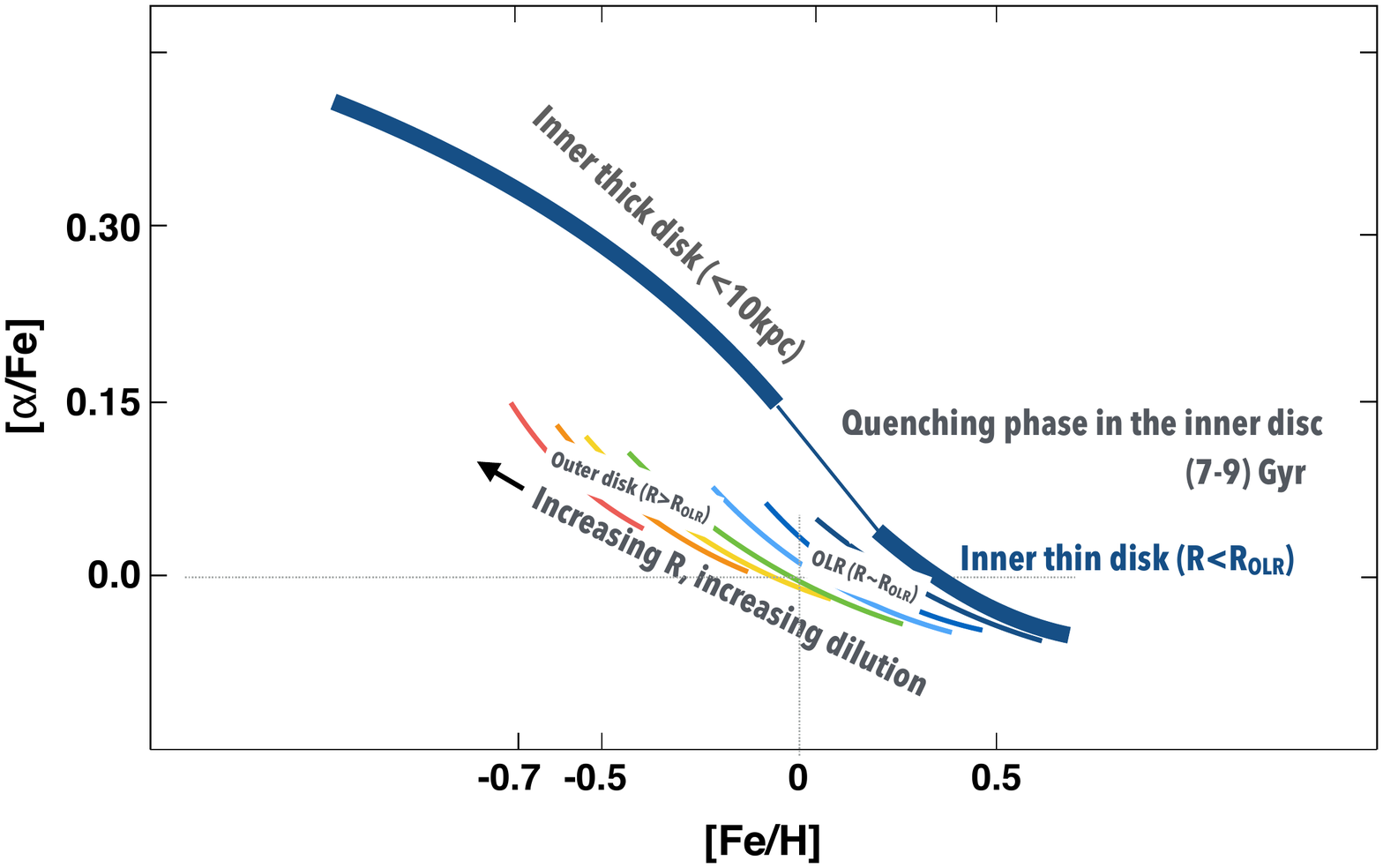}
\includegraphics[trim=30 140 -100 10,clip,width=10.2cm]{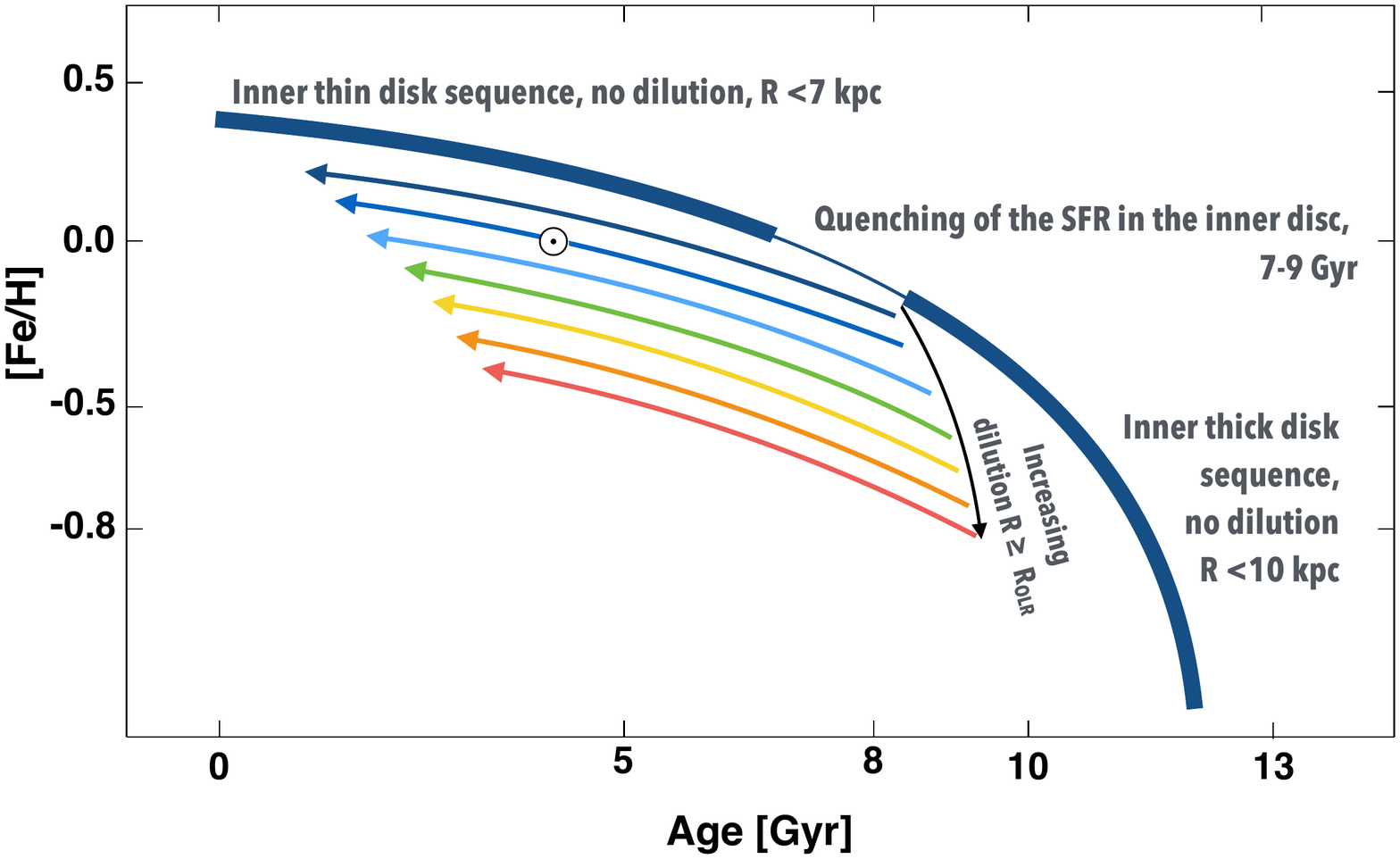}
\includegraphics[trim=30 70 -50 30,clip,width=9.6cm]{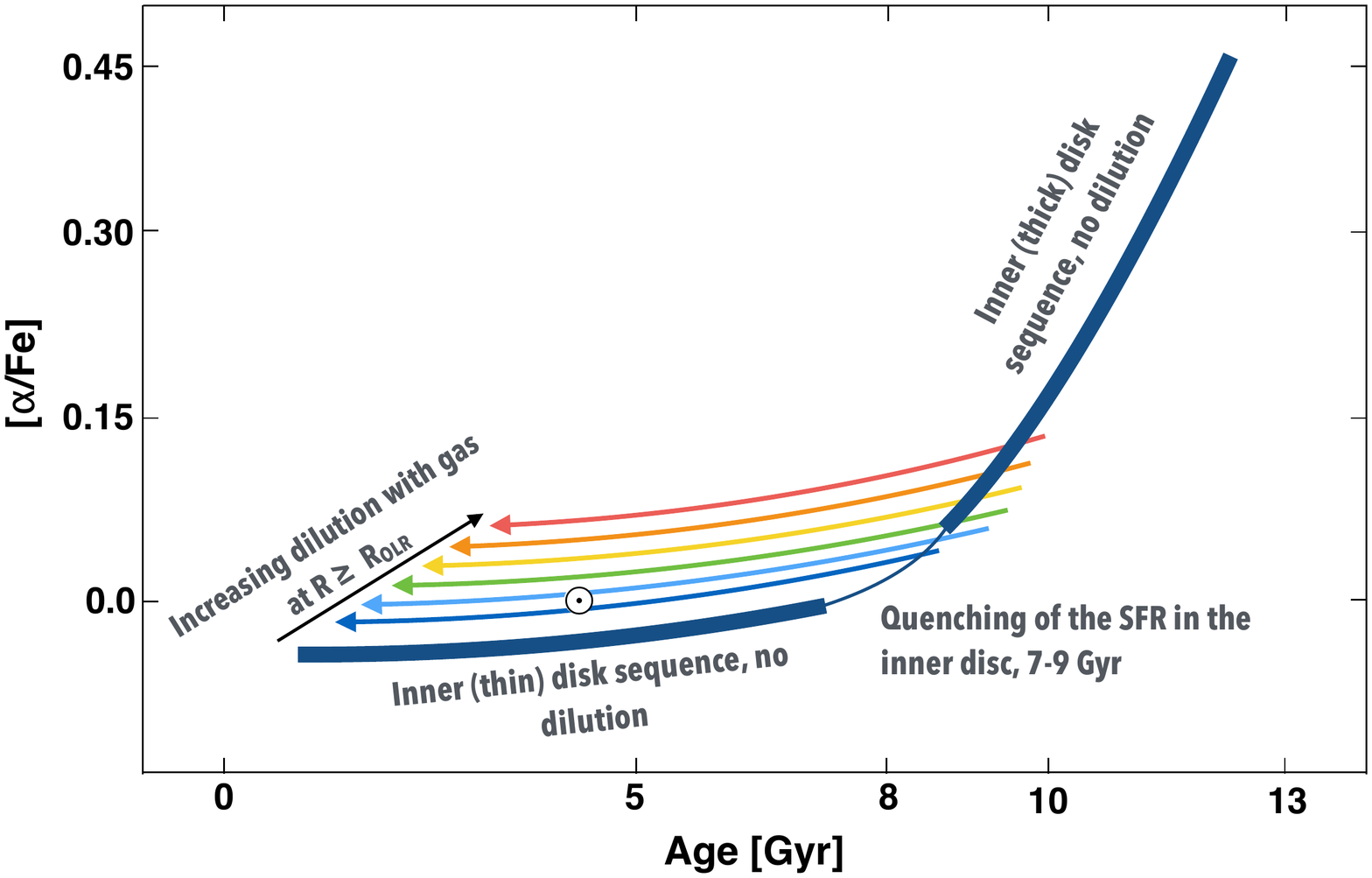}
\caption{Schematic representations of the [Fe/H]-[$\alpha$/Fe],
age-metallicity, and  age-[$\alpha$/Fe] relations according to the analysis
presented in Sects.~\ref{sec:solar} and \ref{sec:outerdisk}. In all
the plots, the thick curve represents the evolution of the inner disk,
which is  described by a closed-box model \citep{haywood2018} with
a two-phase SFH separated by a quenching episode (thinner segment
along this thick curve).  We  expect the evolution
of the inner disk to have a tight, well-defined age-metallicity
relation \citep[see][]{haywood2018}.  At a given epoch, the inner disk,
R$\lesssim$6 kpc, is always the point of maximum metallicity reached
by the Milky Way.  The thinner, colored tracks describe the
evolution of the outer disk at different radii and the initial dilution
increasing with R.  While the dilution seems to have occurred after the
quenching phase at the solar radius, it may have been at earlier times
in the outer disk.}

\label{fig:sketch123}
\end{figure}

The previous models are generalized in the form of three different
plots sketching this evolution in the [Fe/H]-[$\alpha$/Fe], age-[Fe/H],
and age-[$\alpha$/Fe] planes (see Fig.~\ref{fig:sketch123}); they were made 
to understand how these evolutions, generalized to the whole disk, can give rise to
the chemical patterns that we see in the solar vicinity.
We now describe each of these plots and how they represent a generalization of
the scheme that is appropriate for the solar circle.

\begin{itemize}

\item The evolution of the thin disk is sketched in the
[$\alpha$/Fe]-[Fe/H] plane in Fig. \ref{fig:sketch123} (top), represented
by a continuous series of tracks, with about $\sim$0.3 dex increase in
metallicity and about 0.1 dex decrease in [$\alpha$/Fe].  The top thick  blue curve
represents the evolution of the inner disk, with the thick disk phase,
then the quenching episode (thin blue curve), then the evolution
of the thin inner disk. It shows a continuity in the evolution of
 the thick and the thin disks. This evolution
can be reproduced by a chemical evolution model with no dilution,
e.g., by a closed-box model with a two-phase SFH
\citep[see][]{haywood2018}. However, in the outer disks, the situation
is different due to the dilution, and to the likely low star formation
efficiency and lower star formation rate. 
The lower star formation rate will produce fewer metals, and the observed lower star formation
efficiency in outer disks \citep{bigiel2010} also suggests that the production of metals 
will be less efficient in enriching the ISM because of its proportionally larger gas fraction.
 
Due to low star formation rates and efficiencies in the outer disk, we
would expect the chemical evolution tracks to have a smaller range in both
[$\alpha$/Fe] and [Fe/H].  Each colored thin line represents tracks
of the evolution of the outer disk at a given Galactocentric distance,
starting at increasing dilution (decreasing metallicity) with increasing
radii.  Except for the inner thin disk track, which is the
continuation of the thick disk track with no dilution, the other tracks
are thus not connected to the upper $\alpha$-rich sequence, although there is
possibly an indirect dependence through outflows, as discussed in
the previous section. We hypothesized that in fact the initial metallicity
and alpha-enrichment of the outer disk comes from metals formed during
the thick disk phase that polluted it (see Section~\ref{sec:discussion}).
We note that this is different from studies assuming a direct jump
between the two sequences in one continuous chemical evolution \citep[see, e.g.,][]{schonrich2009}.

\item The middle panel of Fig.~\ref{fig:sketch123} sketches the
age-metallicity relation underlying the chemical patterns. Again, the
thick blue curve represents the evolution of the inner disk (roughly
within 6~kpc) and the thin blue line shows that the evolution of the inner disk
is  the continuity in the chemical enrichment of the thick disk after
the quenching episode. Beyond this limit (R$>$ 6~kpc), age-metallicity
relations are diluted with respect to the inner disk evolution. Hence,
at a given age the inner disk evolution is always the most metal rich
of any of the sequences.  In \cite{haywood2018}, we predict that the
age-chemical abundance relations of the inner disk should be very tight.
Before the quenching phase, the main driver of the chemical evolution
of the Milky Way is the formation of the thick disk.  All the curves below
the inner thin disk curve represent the evolution at different radii,
starting at the solar circle and progressing systematically outwards. It
should be noted that the solar vicinity track is not in continuity with
the thick disk evolution. As explained in section \ref{sec:solar},
the solar circle was probably diluted by $\sim$0.2 dex.  The most
metal-poor objects in the outer disk are apparently as old as 9-10 Gyr
(see Fig. \ref{fig:alphafeh-age}, plot a); the tracks at larger radii
start at progressively older ages.

\item The bottom panel of Fig.~\ref{fig:sketch123} sketches  the
corresponding age-[$\alpha$/Fe] relation(s).  Here again, the inner disk
relation is shown as the thick blue  curve.  Since this is a closed-box
with homogeneous evolution, we expect and measured a very tight chemical
evolution, as is seen in particular on the thick disk part of the
relation.  The outer disk evolutionary segments are parallel to the
inner thin disk track, having only slightly higher alpha-abundances
(Fig. \ref{fig:alphafeh-age}) and, for the evolutionary track at the
largest radius (red curve), started forming stars up to 9-10~Gyr ago.

\end{itemize}

In these plots, the red sequence, which corresponds to the evolution
of the far outer disk, the initial metallicity may have been set by the
metals ejected by ouflows at the peak of the star formation rate during
the thick disk phase, while at closer distances to the center of the Milky Way,
the initial metallicity of the gas may have also been contaminated by
the highly enriched gas at the end of the thick disk phase.

In the scenario we have outlined, what we are observing in the stars that
lie at the solar circle is really the superposition of two different
evolutionary sequences. The first is the evolution of the inner disk
(thick blue  curves), which is simple, continuous, and homogeneous over the
scale of the whole inner disk (R$<$6~kpc). The second is the evolution of
the outer disk, which, due to dilution, sets initial conditions when the
star formation commenced and is a function of distance to the Galactic
center. In other words, it is only in the outer disk where the chemical
evolution is distinctly a function of rings of constant radius. This is
the regime where the chemical evolution of the Milky Way is classically
modeled, i.e., as a set of independent rings.

\subsection{Global scenario}

We now summarize the various events that led to the two-phase formation
of the disk.  The following puts together results from various studies
to try to explain the characteristics of the evolution of the disk
\textit{\emph{in toto}}.

(1) The thick disk formed within 3-4 Gyr from 9 to 13 Gyr, in a
starburst phase with a SFR reaching $\sim$12 \Msun/yr in the inner
parts (R$<$10~kpc) of the disk. Feedback and turbulence from the star
formation activity homogenized metals in the thick disk ISM, producing
a flat metallicity gradient. The most intense phase, over the age range
$\sim$10-12~Gyr, generated metal outflows which  polluted the outer disk,
raising the metallicity at $\sim$-0.6 dex.

(2) The velocity dispersion in the gaseous turbulent disk started to decrease 
early, as implied from the observed correlation between the age and stellar
velocity dispersion, where the vertical velocity dispersion 
decreases from more than 40~km~s$^{-1}$ at $\sim$12~Gyr to about 30~km~s$^{-1}$ at 10~Gyr \citep[see ][]{haywood2013}. 
In these conditions, in a less turbulent disk, the bar started to form at an age $\sim$10 Gyr, 
quenched the SFR activity
within the corotation region within $\sim$ 1~Gyr, marking the end of
the thick disk formation. The formation of the OLR at R$\sim$~6kpc
isolated the inner disk from the outside.  Beyond the OLR, the enriched
gas ([Fe/H]$\sim$0 dex) remaining from the thick disk formation mixed
with more metal-poor gas ([Fe/H]$\sim$-0.6 dex) of the outer disk,
establishing a gradient function of the fraction of the metal-rich
and metal-poor gas.

(3) In the inner disk, within the OLR, chemical evolution proceeded
unabated after temporarily quenching, continuously processing the gas
remaining from the formation of the thick disk. \cite{haywood2018} showed
that this evolution can be described with a model with no dilution,
closely approximated for the last 12~Gyr (metallicity above $-$0.7 dex)
by a closed-box model with a break in the star formation between 7 and 9 Gyr (the quenching event).

(4) In the outer disk (R$>$ 6~kpc), chemical evolution continued after
the formation of the thick disk from increasingly lower metallicity gas at larger
R. The gradient is a result of the mixing of the gas polluted in the outer disk with
the gas remaining after the thick disk formation.  The steep gradient
observed up to the radial extent probed in the APOGEE survey (R$\sim$ 15~kpc) shows
that the outer disk can be described as a series of parallel evolutions
that evolved relatively separated from each other.

\section{The Sun as an outer disk star}\label{sec:sun}

We now discuss whether the Sun has the characteristics of an inner or 
outer disk star. Specifically, in this distinction, we mean whether 
or not it formed out of gas that was diluted, which we argue can explain 
the chemical trends within and beyond R$\sim$ 6~kpc.

\begin{figure}
\includegraphics[trim=0 200 0 0,clip,width=11cm]{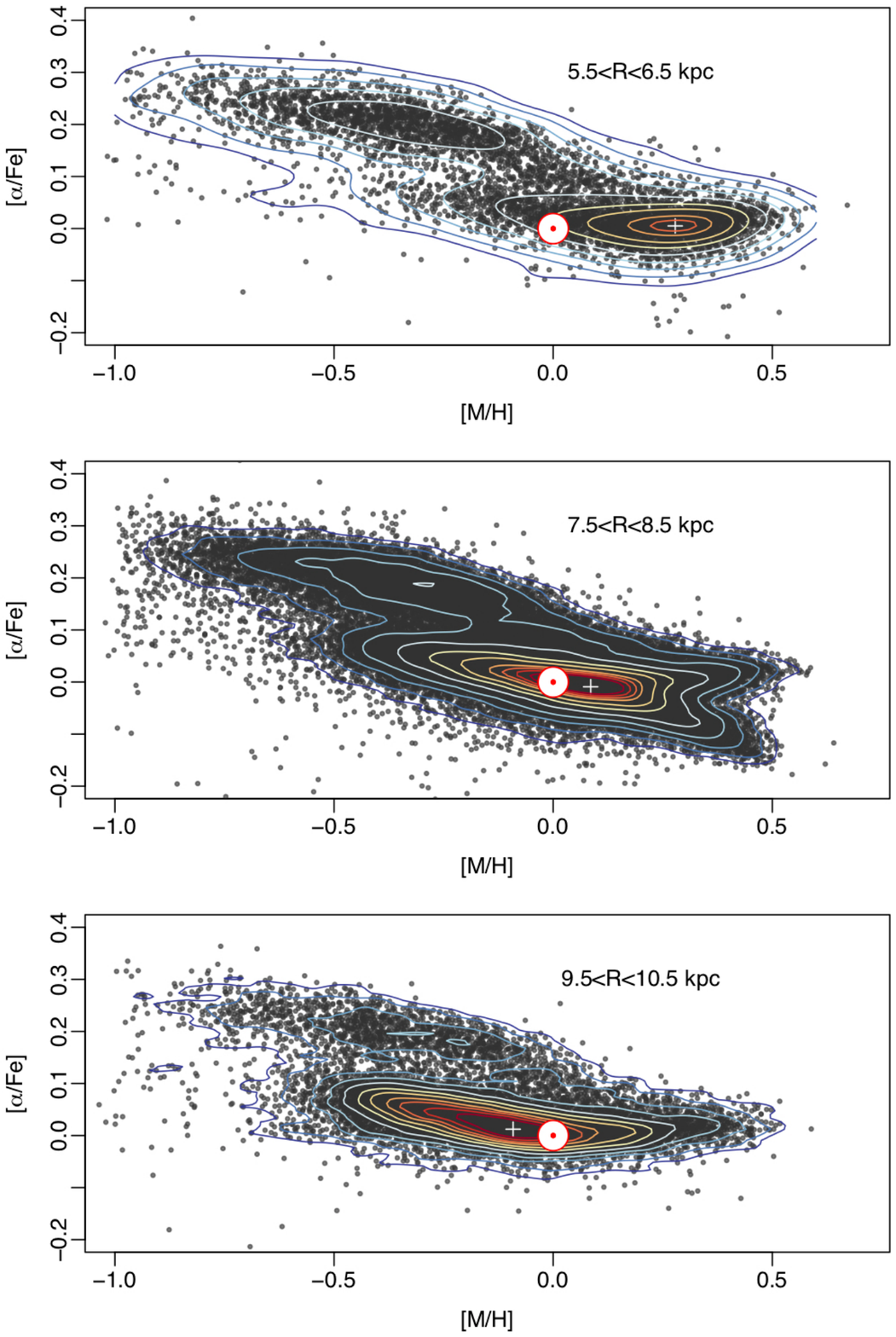}
\caption{[M/H]-[$\alpha$/Fe] distributions of stars in different distance
intervals, 5.5-6.5, 7.5-8.5 and 9.5-10.5 kpc from the Galactic center and
peak metallicity as a function of Galactocentric radius (bottom plot).
These plots illustrate that the region where the probability of finding
a solar metallicity and solar alpha-abundance star is highest beyond
the solar orbit, at R=8.89 kpc. Arguing from chemical offsets that the
Sun comes from the inner disk is not supported by the APOGEE survey.  }
\label{fig:sunpos}
\end{figure}

The Sun is offset compared to the evolution of the inner disk in two
aspects,  in  [$\alpha$/Fe] abundance and in metallicity. This is illustrated by Fig.~\ref{fig:fehgradient} for metallicity, 
and also in Fig.~\ref{fig:sunpos}, which shows the [Fe/H]-[$\alpha$/Fe] distributions
of stars in the APOGEE survey in three different distance intervals, from
5.5 to 10.5~kpc, with density contours, compared to the position of the
Sun.  Inner disk stars at solar metallicity have [$\alpha$/Fe] slightly
above 0.1 dex in APOGEE, as shown on Fig.~\ref{fig:sunpos}. Because
[$\alpha$/Fe] is well correlated with age, it means that the Sun
is too young by a few Gyr compared to inner disk stars of the same
metallicity. Stars on the high-$\alpha$ sequence at solar metallicity have
ages of $\sim$ 9~Gyr, meaning that the Sun is offset by at least 4~Gyr compared
to stars that evolved within the inner disk.  Figure~\ref{fig:sunpos}a
shows that the peak of the low-$\alpha$, inner disk stars (at [Fe/H]
$\sim$ +0.3 and [$\alpha$/Fe] $\sim$ 0) is also separated from
solar metallicity by almost 0.3 dex.  Figure~\ref{fig:fehgradient} is also
a direct indication of the radius at which stars of solar metallicity are the
most common; this is near 9~kpc from the Galactic center.

The whole argument about the Sun having migrated from inner regions is based
on the fact that it would be offset in its chemical properties compared to the population
of the solar vicinity, more akin to the chemistry observed in inner disk stars.
Figure~\ref{fig:sunpos} refutes this, and it is
certainly an outlier to the inner disk chemistry.  If we are to follow
strictly the indication provided by Fig.~\ref{fig:fehgradient}, then the
Sun has a higher probability of originating from $\sim$8.9 kpc, 
being slightly offset (by 0.07 dex) in metallicity
compared to populations at R=8kpc.

Our position is supported 
by \cite{martinez2015} who found, by integrating the orbit of the Sun
backwards, that for all their assumed bar+spirals potentials, the origin
of the Sun is always in the outer disk, sometimes as far as 11~kpc from the Galactic center. In
the case where \cite{martinez2015} assume a bar with a pattern speed of
42 km.s$^{-1}$.kpc$^{-1}$ (OLR of the bar between 9 and 10kpc, see their
Figure 4, which could be the position of the OLR today, see references given above) and weak spiral arms, they find that the Sun could, at most,
have migrated from the {\it \emph{outer}} disk by about -0.83 kpc, which means,
assuming R$_{\odot}$=8 kpc, that the Sun originated from R=8.83 kpc. This is
in excellent agreement with the estimate given by the metallicity
gradient above.  In all the cases they considered, \cite{martinez2015}
found that the Sun migrated from the outer disk by small distances.
This points to the result that, both for its chemistry and its dynamics, the
Sun is not an inner disk object, at variance with what has been found in
the last 20 years \citep{wielen1996,minchev2013, kubryk2015,frankel2018}.  On the
contrary, we would argue that the distributions studied here show that the
Sun is much more compatible with chemical evolution with dilution, which
we believe characterizes the evolution beyond $\sim$ 6~kpc. We conclude
the Sun is an outer disk star, and has a higher probability of originating from R$>$8 kpc
than the contrary.

\section{Discussion}\label{sec:discussion}

\subsection{Previous interpretations}

The picture we propose is different from the standard gas infall schemes in the following ways:\\

{\it Thick disk growth.} The initial growth of the metal content 
of the ISM is explained by a massive population of stars which we associate with the thick disk.
Even though this population is only 15-25\% of the local
surface density, it represents about half the overall stellar mass of the Milky Way \citep[see][]{snaith2015,haywood2016a}. 
We argue that the high level of turbulence and feedback in the ISM at the epoch of thick disk formation allowed
the outskirts of the thick disk to have a chemical evolution similar to the inner regions.
This position is supported by both the observation of the Milky Way's metallicity gradient, 
the gradient of galaxies at the epoch of thick disk formation, and in galaxy simulations. 
Hence, the disk at the Sun radius was enriched by a massive stellar population of $\sim$2.10$^{10}$ \Msun. 
In this scenario, the thick disk is not formed inside-out, and the long timescale, radially dependent 
accretion of gas is not necessary;  the chemical evolution is   described well by a closed-box 
\cite[with some specific SFH, see][]{snaith2015,haywood2018}.
In inside-out scenarios, the enrichment occurs independently in rings at a particular radii: 
the outer regions have low gas surface densities and a slower enrichment, 
thus metallicities reached at the solar ring at the end 
of the thick disk phase are low (usually in the range -1.0$<$[Fe/H]$<$-0.5 dex).
This obviously is insufficient to provide the level of enrichment observed (see, 
e.g., \cite{chiappini1997,colavitti2009,minchev2013}).
In models including radial migration, thick disk stars formed in the inner galaxy (and  therefore 
more metal rich) are allowed to reach the solar circle. 
For instance, in \cite{kubryk2015}, the thick disk at the end of its evolution has a metallicity of $\sim$-0.8 dex
at the solar radius, but solar metallicity thick disk stars are allowed to migrate to solar vicinity.
These models, however, predict a spread in metallicity and [$\alpha$/Fe] abundance of about 0.2 dex
at the end of the thick disk phase.
This is excluded by observations (Fig. \ref{fig:alphafeh}, bottom plot). The combination of inside-out scenario and radial migration is simply not 
compatible with the observations (see \cite{haywood2015} for more details and other arguments against an inside-out
thick disk formation). 
\\

 {\it Pre-enrichment and dilution.} The second phase explains the formation of the outer thin disk (R$>$6 kpc) to which our solar neighborhood belongs: an inflow of gas   dilutes the ISM left by the thick disk phase to a metallicity compatible with the metallicity of the oldest thin disk stars (about -0.2 dex). The inflow of gas actually plays  the opposite role in 
standard chemical evolution models, where the slow infall is used to limit as much as possible 
the dilution of metals.

Thus, in our scenario, the inflow of gas is not invoked to control the width of the MDF, although it 
contributes to determining the metallicity of the gas from which thin disk stars were formed, 
setting the initial metallicity of the outer thin disk decreasing with radius.

Pre-enrichment of the disk  (or prompt initial enrichment, PIE) has been proposed in the past whereby top-heavy IMF 
provides enough metals from massive stars to solve the G-dwarf problem (see \cite{truran1971}).
A model with thick disk pre-enrichment has also been proposed by \cite{gilmore1986}, 
who derived a mass ratio between the two disks of 1/4, which was sufficient assuming that the thick disk 
would pre-enrich the gas to [Fe/H]$\sim$-0.6. However, this is no longer  compatible  with 
the solar metallicity of the youngest thick disk stars, and pre-enrichment to [Fe/H]=0, as is 
now required, would not be possible to reach if the thick-to-thin disk mass ratio was only 1/4.

In 2001, Pagel proposed an interesting interpretation of the then recent discovery of the two separate sequences
of the thin and thick disks in [Fe/H]-[$\alpha$/Fe]. He pointed out that a combination of pre-enrichment by the 
thick disk and inflow at the beginning of the thin disk phase was necessary to explain the thin disk at the solar vicinity, 
and his Fig.~3 foreshadows our Fig.~\ref{fig:sketch123}.

\cite{haywood2001} found that the solar vicinity data could be compatible with a closed-box model, 
provided that the thick disk contribution perpendicular to the Galactic plane was taken into account, which
is usually not done \citep{sommer1991}.
 The closed-box model envisaged in \cite{haywood2001} was similar to the one developed in \cite{snaith2015}, except 
that a constant SFR was assumed, producing a less significant metal-poor star tail than the bimodal MDF now 
observed in the inner disk, and which requires a more active SFR in the first Gyrs, during the thick disk phase.
Observations at that time did not show that the solar vicinity had peculiar chemical trends and 
that the thick disk had reached solar vicinity. Hence, the models could be fitted to the local MDF 
(with scale height 
corrections), but would not be adequate to describe the inner disk MDF that we now know due to the APOGEE survey.

 {\it Origin of the low-$\alpha$ sequence.} \cite{nidever2014} discuss the possibility that the low-$\alpha$ sequence could result from 
an evolution at low star formation efficiency and significant outflows occurring in the outer disk.
Their Fig. 16 reproduces two such possible evolutionary tracks. A single such sequence would not be able
to reproduce the complicated age-chemical structure of Fig. \ref{fig:alphafeh-age}, but a series of them, produced 
by varying the star formation efficiency, could; however, 
this option  presents two inconveniences.
First, as noted by \cite{nidever2014}, the progenitor stars of the low-$\alpha$ sequence are not seen in the outer disk. 
These would not be expected to exist only at metallicities lower than the tail of the low-$\alpha$ sequence (or [Fe/H]$<$-0.6), but since the oldest stars of the low-$\alpha$ sequence cover the whole range of metallicities, 
we would expect these progenitor stars to be present at all metallicities. They are not observed.
Second, the solar vicinity data analyzed here demonstrate that the initial metallicity of the thin disk at the solar radius 
was set by dilution.
Given the continuity observed on the gradient of Fig. \ref{fig:fehgradient}, it is difficult to think that this scheme
would be valid only at the solar radius. We note, however, that an evolution at low star formation efficiency, as described in \cite{nidever2014}, arising from more pristine gas, unpolluted by thick disk metals could have occurred at larger distances than those probed by APOGEE.

\subsection{What does the G-dwarf distribution tells us?}

What is the meaning of the G-dwarf metallicity distribution, as a constraint for chemical evolution, 
in this new context?
The stars that comprise the local MDF are mostly  younger
than 7 Gyr (80\% in our sample) and have metallicities above -0.2 dex (73\%). 
These stars were born from a mixture that can only be found approximately at  the solar ring.
In our scenario, this mixture was made from inner disk gas which was enriched through a closed-box-type 
evolution and outer disk gas also pre-enriched to $\sim$-0.6 dex, as described in sections \ref{sec:solar} and \ref{sec:outerdisk}.

In infall models, the width of the local MDF is used to constrain the infall timescale at the solar radius. 
The wider the MDF, the smaller the accretion timescale.
In our scenario, the gaseous mixture from which these stars were born was in place before their
formation, the necessary enrichment being provided by the formation and evolution of the thick disk. 
The G dwarfs responsible for the 
enrichment seen at the solar vicinity are therefore not missing, they are only not present at the solar 
vicinity in proportion relative to their effect on chemical evolution, because chemical evolution cannot 
be modeled as a strictly local process.
In our view, the width of the local MDF is therefore entirely determined by an initial enrichment that was set by 
a global process in the Milky Way (the formation of the thick disk)  
and subsequent SFH. It is not a measure of the infall timescale at the solar radius.

\subsection{``The fault, dear Brutus, is not in our stars...''}

It is the mixing of stars in the disk that allows us to sample stars
at solar radius that dominate at other radii and which, together with
the extension of spectroscopic surveys well beyond the solar radius,
 allow for new insights into the chemical patterns arising from
the chemical evolution of the disk. In turn, how does this mixing
affect our conclusions? Mixing arises from the secular increase in
the random motion of stars and their kinetic energy and/or from a
change in their angular momenta, often dubbed blurring and churning
in the literature.  Blurring, by increasing the radial excursions of
stars, contaminates other radii and increases the observed metallicity
dispersion at a given radius. Because our measured gradient is based
on the metallicity of the peak of the distribution, it is unlikely
to be significantly affected by the increase in the metallicity
dispersion at a given radius due to blurring, which is usually thought
to be modest \citep[see][]{binney2007,schonrich2009,hayden2015}.
The effect of churning could be more important. As mentioned
previously, the redistribution of angular momentum by the bar has the
effect of moving material, both stars and gas, from the inner parts
of galaxies to the OLR (stars: \cite{halle2015, halle2018}; gas:
\cite{simkin1980,rautiainen2000}).  It is not surprising, in these
circumstances, that there is no metallicity gradient for stars within 6
kpc. First because the thick disk left no gradient and second because
the action of the bar redistributes metals throughout the inner
disk. These effects explain why so many metal-rich stars are found up to R$\sim$ 6
kpc.  It was shown in \cite{halle2015, halle2018} that this redistribution
stops at the OLR, beyond which stars are not allowed to migrate
via churning. This is also the case for the gas, which, accumulating at
the OLR, has a tendency to form rings \citep{simkin1980,rautiainen2000}.
Therefore, by moving the enriched gas from the corotation to the OLR,
the formation of the bar may have provided fuel to form metal-rich stars
even very near the solar orbit. In this respect, the redistribution of
gas may be more important than radial migration of stars to explain the
amount of metal-rich stars found up to R$\sim$6~kpc.

The fact that the metallicity of the thin disk shows a steep gradient beyond 6 kpc (Fig. \ref{fig:fehgradient}) 
supports the idea that the metal-rich material cannot go on their guiding radius beyond the OLR in any significant 
number, as shown in \cite{halle2015}, while stars that are far from their initial guiding radius are more likely to be there
because of blurring effects \citep[see][]{halle2018}.
It is very  possible, if the Milky Way bar is long-lived and therefore the OLR maintains its barrier effect 
(although it will shift to larger radii), that the only metal-rich fuel that has been 
available to the outer disk is the one provided by the formation of the thick disk at solar metallicity.

The steep gradient observed in Fig.~\ref{fig:fehgradient} supports the idea that radial
migration had at most a minor role in redistributing stars at the solar vicinity. 
The gradient at R$>$6~kpc shows that stars of a given metallicity are strongly dominating at the 
radius indicated by the gradient, and  essentially only a small fraction are  seen at other places in the disk of the Milky Way. 
This is supported by other studies of the solar vicinity.
For example, the results of \cite{hayden2018} are illustrative. Out of their original 
2364 stars in their sample from the GES survey, 51 have [Fe/H]$>$0.1 dex and perigalacticon $>$ 7 kpc 
and are likely to be migrators (assuming that stars on the most circular orbits are more likely
to migrate, and among them, stars with the most extreme metallicities) or 2\% of the stars. 
Even so, one may find 
these criteria to be generous because the solar neighborhood is likely to form stars 
with [Fe/H]$\sim$ 0.1 dex, and because of the errors in the metallicities and the shape of the MDF, 
many more stars are likely to have estimated metallicities above 0.1 dex than below 0.1 dex. 
Raising the limit in metallicity to 0.25 dex, \cite{hayden2018} find seven stars in their sample 
that have a perigalacticon $>$ 7 kpc, or 0.3\%. These are very likely to be real migrators 
and are important for explaining the spread in metallicity at a given radius. Nonetheless, it is a very 
small fraction, and it is difficult to argue on these grounds that churning has affected
a significant number of stars in the solar vicinity and that it could affect our overall conclusions. 

\section{Conclusions}\label{sec:conclusion}

We find that the disk chemical evolution has followed two different paths depending on  the distance from the Galactic center where 
the stars originated. One corresponds 
to the evolution of the inner disk, and is  described well by a model where most of the gas 
was accreted early and evolved homogeneously, technically  approximated by the closed-box 
model described in \cite{haywood2018}, with a two-phase SFH determined in \cite{snaith2014, snaith2015}.
This evolution is valid up to $\sim$ 6~kpc from the Galactic center and is what defines the inner disk.
The formation of the outer disk would arise from the gas left by the formation 
of the thick disk at solar metallicity, mixed with more metal-poor gas, in a ratio which is function of R.
The main points of this scenario are as follows:

\begin{itemize}

\item Due to the vigorous star formation during the formation and evolution of the thick disk inducing high
turbulence and its concomitant strong gas phase mixing, the entire disk to R$\sim$10~kpc  -- before the formation of the bar and the OLR at around 6~kpc -- was enriched 
due to this population. Hence, regions like the 
solar ring, at the periphery of the thick disk, benefited from the enrichment of an entire massive population, although
it represents only a small fraction of the surface density of the disk at the solar vicinity today.

\item The thick disk enriched the disk to solar metallicity. An additional supply of more metal-poor gas must 
then have been available to dilute the ISM to -0.2 dex (the initial metallicity of the thin
disk at the solar vicinity). Combined with the gas left by the thick disk, it provided the fuel necessary 
to form the thin disk. 

\item The gas present in the outer disk must have had a metallicity of 
about -0.6 dex at the time the thin disk started to form, based on what we can measure on the oldest outer thin disk stars. 
This gas is a good candidate for the dilution of the gas left by the thick disk phase.
At the solar ring, the metallicity of the outer disk gas ($\sim$-0.6 dex) imposes that it contributed to two-thirds of the ISM
present at the end of the thick disk formation. The other one-third corresponds to the gas left over from the formation of the thick disk. Although there is no clue to the origin of the chemical composition of the gas of 
the outer disk, a possibility is that it may have been pristine gas polluted by outflows generated during 
the formation of the thick disk \citep{haywood2013, lehnert2014}.

\item We suggest that the mixing of the gas left from the thick disk formation
with more pristine gas from the outer disk possibly occurred at the epoch of the formation of the bar 
and the establishment of the OLR at about R$\sim$6 kpc. Detailed simulations are needed to test this hypothesis.

\item The decreasing fraction of gas left over from the formation of the thick disk induced a negative metallicity gradient in the disk at R$>$6 kpc.

\item From the chemical evolution point of view, the Sun is not an inner disk star, but is well on the path 
of chemical evolution with dilution. Thus, it is better described as an outer disk than an inner disk object, 
as is also supported by its orbital properties (see \cite{martinez2015}).
The Sun is typical of the stars present at solar vicinity and does not seem
to have any of the properties of the inner disk objects.

\item If this scenario is correct, it means that the local G-dwarf metallicity distribution has no connection 
with the infall history of our Galaxy (but the inner disk MDF has; see \citet{haywood2018}), and therefore cannot be used as evidence of long-timescale gas accretion. 
The metallicity distribution of the solar vicinity is simply the result of a disk of gas pre-enriched to an 
initial metallicity of -0.2 dex and a mean SFR of about 1-3\Msun/yr, with no prolonged infall of gas. 

\end{itemize}

\begin{acknowledgements}
The Agence Nationale de la Recherche (ANR) is acknowledged for its financial support through the MOD4Gaia project (ANR-15-CE31-0007, P. I.: P. Di Matteo), and also for providing the postdoctoral grant for Sergey Khoperskov.
The authors are grateful to the referee for their comments and suggestions. 
Funding for the Sloan Digital Sky Survey IV has been provided by the Alfred P. Sloan Foundation, the U.S. Department of Energy Office of Science, and the Participating Institutions. SDSS acknowledges support and resources from the Center for High-Performance Computing at the University of Utah. The SDSS web site is www.sdss.org.

SDSS is managed by the Astrophysical Research Consortium for the Participating Institutions of the SDSS Collaboration including the Brazilian Participation Group, the Carnegie Institution for Science, Carnegie Mellon University, the Chilean Participation Group, the French Participation Group, Harvard-Smithsonian Center for Astrophysics, Instituto de Astrofísica de Canarias, The Johns Hopkins University, Kavli Institute for the Physics and Mathematics of the Universe (IPMU) / University of Tokyo, Lawrence Berkeley National Laboratory, Leibniz Institut für Astrophysik Potsdam (AIP), Max-Planck-Institut für Astronomie (MPIA Heidelberg), Max-Planck-Institut für Astrophysik (MPA Garching), Max-Planck-Institut für Extraterrestrische Physik (MPE), National Astronomical Observatories of China, New Mexico State University, New York University, University of Notre Dame, Observatório Nacional / MCTI, The Ohio State University, Pennsylvania State University, Shanghai Astronomical Observatory, United Kingdom Participation Group, Universidad Nacional Autónoma de México, University of Arizona, University of Colorado Boulder, University of Oxford, University of Portsmouth, University of Utah, University of Virginia, University of Washington, University of Wisconsin, Vanderbilt University, and Yale University.
\end{acknowledgements}

\clearpage

\end{document}